\documentclass[draft,onecolumn]{IEEEtran}
\usepackage{amsmath,bm}
\usepackage{amsfonts}
\usepackage{amssymb}
\usepackage[final]{graphicx}
\usepackage{mathtools}
\usepackage{upgreek}
\usepackage[left=2cm,right=2cm,top=2cm,bottom=2cm]{geometry}
\usepackage{algorithm,algorithmic}
\usepackage{paralist}
\usepackage[hidelinks,colorlinks=false]{hyperref}
\usepackage{tikz,pgfplots}
\usepackage{multicol}
\usepackage{subfigure}
\newtheorem{theorem}{Theorem}

\newtheorem{proposition}{Proposition}

\newtheorem{definition}{Definition} 
\usepackage{amsmath}
\usepackage{breakurl}
\usepackage{amssymb}
\newcommand{\Tpiy}{T(\pi,y)}
\newcommand{\Spiy}{\sigma(\pi,y)}

\newcommand{\Root}{X}
\newcommand{\Obs}{Y}
\newcommand{\prob}{\mathbb{P}}
\newcommand{\state}{X}
\newcommand{\Time}{t}
\newcommand{\statedim}{S}

\newcommand{\statespace}{\mathcal{S}}
\newcommand{\tp}{P}
\newcommand{\belief}{\pi}
\newcommand{\Belief}{\Pi}
\newcommand{\action}{u}
\newcommand{\obspace}{\mathcal{Y}}
\newcommand{\obp}{B}
\newcommand{\filter}{T}
\newcommand{\history}{Z}
\newcommand{\E}{\mathbb{E}}
\newcommand{\obs}{Y}
\newcommand{\filternorm}{\sigma}
\newcommand{\one}{\textbf{1}}
\newcommand{\Real}{\mathbb{R}}
\newcommand{\policy}{\mu}
\newcommand{\policyspace}{\mathcal{U}}
\newcommand{\actionspace}{\mathcal{A}}
\newcommand{\discount}{\rho}
\newcommand{\discountedreward}{J}
\newcommand{\argmax}{\operatornamewithlimits{argmax}}
\newcommand{\optpolicy}{\policy^*}

\newcommand{\valuefunction}{V}

\newcommand{\looks}{l}

\newcommand{\g}{g}
\newcommand{\veng}{viewer engagement }
\newcommand{\reals}{{\rm I\hspace{-.07cm}R}}
\usepackage{footmisc}
\newcommand{\reward}[1] {r^\prime #1}
\graphicspath{{./figs/}}
\usepackage{cleveref}
\crefformat{equation}{(#2#1#3)}
\crefrangeformat{equation}{(#3#1#4) to~(#5#2#6)}
\crefmultiformat{equation}{(#2#1#3)}%
{ and~(#2#1#3)}{, (#2#1#3)}{ and~(#2#1#3)}
\usepackage{enumitem}
\newcommand\given[1][]{\:#1\vert\:}
\begin{document}
\title{Opportunistic Advertisement Scheduling in Live Social Media: A Multiple Stopping Time POMDP Approach}
\author{%
\IEEEauthorblockN{Vikram Krishnamurthy \\}
\IEEEauthorblockA{Department of Electrical and Computer Engineering,\\ Cornell University, Ithaca, NY \\\textit{vikramk@cornell.edu} \\} 
\IEEEauthorblockN{Anup Aprem \\}
\IEEEauthorblockA{Department of Electrical and Computer Engineering,\\ University of British Columbia, Vancouver, BC, Canada \\ \textit{aaprem@ece.ubc.ca} \\}
\IEEEauthorblockN{Sujay Bhatt \\}
\IEEEauthorblockA{Department of Electrical and Computer Engineering,\\ Cornell University, Ithaca, NY \\ \textit{sh2376@cornell.edu}\\}
} 

\IEEEtitleabstractindextext{%
\begin{abstract}
	Live online social broadcasting services like YouTube Live and Twitch have steadily gained popularity due to improved bandwidth, ease of generating content and the ability to earn revenue on the generated content.  
	In contrast to traditional cable television, revenue in online services is generated solely through advertisements, and depends on the number of clicks generated. 
	Channel owners aim to opportunistically schedule advertisements so as to generate maximum revenue. 
	This paper considers the problem of optimal scheduling of advertisements in live online social media. 
	The problem is formulated as a multiple stopping problem and is addressed in a partially observed Markov decision process (POMDP) framework. Structural results are provided on the optimal advertisement scheduling policy. By exploiting the structure of the optimal policy, best linear thresholds are computed using stochastic approximation. 
	The proposed model and framework are validated on real datasets, and the following observations are made: (i) The policy obtained by the multiple stopping problem can be used to detect changes in ground truth from online search data (ii) Numerical results show a significant improvement in the expected revenue by opportunistically scheduling the advertisements. The revenue can be improved by $20-30\%$ in comparison to currently employed periodic scheduling. 
\end{abstract}
\begin{IEEEkeywords}
scheduling, advertisement, live channel, POMDP, multiple stopping, structural result
\end{IEEEkeywords}}

\maketitle

\section{Introduction}
Popularity of online video streaming has seen a sharp growth due to improved bandwidth for streaming and the ease of sharing User-Generated-Content (UGC) on the internet platforms. One of the primary motivations for users to generate content is that platforms like YouTube, Twitch etc., allow users to generate revenue through advertising and royalties. The revenue of Twitch which deals with live video gaming, play through of video games, and e-sport competitions, is around 3.8 billion for the year 2015, out of which 77\% of the revenue was generated from advertisements\footnote{\url{http://www.tubefilter.com/2015/07/10/twitch-global-gaming-content-revenue-3-billion/}}.    

Some of the common ways advertisements (ads) are scheduled on pre-recorded video contents on social media like YouTube are pre-roll, mid-roll and post-roll; where the names indicate the time at which the ads are displayed. In a recent research on viewer engagement, Adobe Research\footnote{\url{https://gigaom.com/2012/04/16/adobe-ad-research/}} concluded that mid-roll video ads constitute the most engaging ad type for pre-recorded video contents, outperforming pre-roll and post-rolls when it comes to completion rate (the probability that the ad will not be skipped). Viewers are more likely to engage with an ad if they are interested in the content of the video that the ad is been inserted into. Most content sharing platforms hosted {\em pre-recorded videos}, until recently, owing to higher bandwidth requirements of real-time video streaming. However, this has changed lately with improved bandwidths (e.g.\ Google Fiber, Comcast Xfinity) and well known content sharing websites like YouTube and Facebook making provisions for {\em live streaming videos} to capture major events in real time\footnote{For example, as early as 2011, millions watched the Royal wedding live through YouTube\footnotemark, and, more recently, the Caribbean Premier League is scheduled to be broadcast live on Facebook\footnotemark}. Online live video streaming, popularly known as ``Social TV", now boasts a number of popular applications like Twitch, YouTube Live, Facebook Live, etc. 

When a channel is streaming a \textit{live video}, the mid-roll ads need to be scheduled manually\footnotemark. Twitch allows only periodic ad scheduling~\cite{SOW13} and YouTube and other live services currently offers no automated method of scheduling ads for live channels.  
The ad revenue in live channel depends on the click rate (the probability that the ad will be clicked), which in turn depend on the viewer engagement with the channel content. 
Hence, ads need to be scheduled when the viewer engagement of the channel is high. 
The problem of optimal scheduling of ads has been well studied in the context of advertising in television; see~\cite{BBM04},\cite{PC15}, \cite{SSS15} and the references therein. However, scheduling ads on \textit{live} online social media is different from scheduling ads on television in two significant ways~\cite{KM11}: (i) real-time measurement of viewer engagement (ii) revenue is based on ads rather than a negotiated contract.  Prior literature on scheduling ads on social media is limited to ad scheduling in real-time for social network games, where the ads are served to either the video game consoles in real time over the Internet~\cite{TC13}, or in digital games that are played via major social networks~\cite{TST11}.  
\footnotetext{{\url{http://www.telegraph.co.uk/news/uknews/royal-wedding/8460801/Royal-wedding-Kate-and-William-marriage-live-on-YouTube.html}}} \footnotetext{{\url{http://www.si.com/tech-media/2016/07/06/caribbean-premier-league-matches-facebook-live}}} \footnotetext{{YouTube Live: Slate and Ad Insertion~\url{https://is.gd/i6c7ku}}}

{\em Problem Formulation: }This paper deals with optimal scheduling of ads on live channels in social media, by considering viewer engagement, termed as \emph{active scheduling}, to maximize the revenue generated from the advertisements. We model the \veng of the channel using a Markov chain~\cite{AMM12,AMM10}. The \veng of the content is not observed directly, however, noisy observation of the \veng is obtained by the current number of viewers of the channel. Hence, the problem of computing the optimal policy of scheduling ads on live channel can then be formulated as an instance of a stochastic control problem called the \emph{partially observable Markov decision process} (POMDP).  To the best of our knowledge, this is the first time in the literature that ad scheduling in live channels on social media is studied in a POMDP framework.  
The main contribution of this paper is two-fold:
\begin{itemize}
	\item[1.)] We provide a POMDP framework for the optimal ad-scheduling problem on live channels and show that it is an instance of the \emph{optimal multiple stopping problem}. 
	\item[2.)] We provide structural results of the optimal policy of the multiple stopping problem and using stochastic approximation compute the best approximate policy. 
\end{itemize} 

{\em Structural Results: }The problem of optimal multiple stopping has been well studied in the literature see~\cite{nak85}, \cite{sta87}, \cite{Nik99}, \cite{Ann15} and the references therein. 
The optimal multiple stopping problem generalizes the classical (single) stopping problem, where the objective is to stop once to obtain maximum reward.  
Nakai~\cite{nak85} considers optimal $L$-stopping over a finite horizon of length $N$ in a partially observed Markov chain. 
More recently, \cite{Ann15} considers $L$-stopping over a random horizon. 
The state of the finite horizon partially observed Markov chain in~\cite{nak85} above can be summarized by the ``belief state''\footnote{The belief state is a sufficient statistic for all the past observations and actions~\cite{bertsekas1995dynamic}.}.  
For a stopping time POMDP, the policy can be characterized by stopping region (set of belief state where we stop) and continuance region (set of belief states where we do not stop). 
Nakai~\cite{nak85} shows that there are $N \times L$ stopping regions corresponding to each time index and stops and these regions form a nested structure. 
However, in live channels, the time horizon $N$ is very large (in comparison to decision epochs) or initially unknown. Therefore, we extend the results in Nakai~\cite{nak85} to the infinite horizon case. 
The extension is both important and non-trivial. 
In the infinite horizon case, the policy is stationary (the stopping regions do not depend on the time index) and hence $L$ stopping regions characterize the policy. 
We obtain similar structural results as~\cite{nak85} in the infinite horizon case.  

Our main structural result (Theorem~\ref{thm:main}) is that the optimal policy is characterized by a switching curve on the unit simplex of Bayesian posteriors (belief states). To prove this result we use the monotone likelihood ratio stochastic order since it is preserved under conditional expectations. However, determining the optimal policy is non-trivial since the policy can only be characterized on a partially ordered set (more generally a lattice) within the unit simplex. We modify the MLR stochastic order to operate on line segments within the unit simplex of posterior distributions. Such line segments form chains (totally ordered subsets of a partially ordered set) and permit us to prove that the optimal decision policy has a threshold structure. Having established the existence of a threshold curve, Theorem~\ref{thm:coefficent:MLR:condition} and Theorem~\ref{thm:constraints:parameter} gives necessary and sufficient conditions for the best linear hyperplane approximation to this curve. Then a simulation-based stochastic approximation algorithm (Algorithm~\ref{algo:policy:gradient:algorithm}) is presented to compute this best linear hyperplane approximation. 

{\em Context: }The optimal multiple stopping problem can be contrasted to the recent work on sampling with ``causality constraints''. In sampling with causality constraints, not all the observations are observable. \cite{ER15} considers the case where an agent is limited to a finite number of observations (sampling constraints) and must adaptively decide the observation strategy so as to perform quickest detection on a data stream. The extension to the case where the sampling constraints are replenished randomly is considered in~\cite{GBL14}. In the multiple stopping problem, considered in this paper, there is no constraint on the observations and the objective is to stop $L$ times at states that correspond to maximum reward. 

The optimal multiple stopping problem, considered in this paper, is similar to sequential hypothesis testing~\cite{Lai97,Lai01}, sequential scheduling problem with uncertainty~\cite{NJ10} and the optimal search problem considered in the literature. \cite{ST05} and \cite{LPVZ15} consider the problem of finding the optimal launch times for a firm under strategic consumers and competition from other firms to maximize profit. However, in this paper we deal with sequential scheduling in a partially observed case.  \cite{WRA11},\cite{AKL16} consider an optimal search problem where the searcher receives imperfect information on a (static) target location and decides optimally to search or interdict by solving a classical optimal stopping problem ($L=1$). However, the multiple-stopping problem considered in this paper is equivalent to a search problem where the underlying process is evolving (Markovian) and the searcher needs to optimally stop $L>1$ times to achieve a specific objective. 

The paper is organized as follows: 
Section~\ref{sec:system:model} provides a model of a live channel and introduces the notations, assumptions and key definitions. 
Section~\ref{sec:scheduling:policy} provides the main results of the paper. 
First, similar to~\cite{nak85}, we show that the stopping regions of the optimal ad scheduling policy form a nested structure. 
Second, we show the threshold structure of the optimal ad scheduling policy.  
In Section~\ref{sec:spsa:MLR:threshold}, we use the nested property of the stopping regions and the threshold property in Section~\ref{sec:scheduling:policy} and stochastic approximation algorithm to compute the best approximate policies using linear thresholds. Such linear threshold policies are computationally inexpensive to implement. 
In Section~\ref{sec:numerical:results}, we validate the model on three different datasets. 
First, we illustrate the analysis using a synthetic dataset and verify the performance of the optimal ad scheduling policy against conventional scheduling policies. 
Second, we show that the policy obtain by the multiple stopping problem can be used to detect changes in ground truth using data from online search. 
Finally, we use real datasets from YouTube Live and Twitch to optimally schedule multiple ads ($L > 1$) in a sequential manner so as to maximize the revenue.  
\section{Opportunistic Scheduling on Live Channels: Model and Problem Formulation}
\label{sec:system:model}
\subsection{Live Channel Model}
In this section, we develop a model of the live channel. 
The three main components of the live channel are  
\begin{inparaenum}[i)]
\item Viewer engagement: How to model the \veng of a live channel?
\item Dynamics of the channel viewers: How does the number of viewers vary with respect to the engagement?
\item Reward of the channel owner: What is the (monetary) reward obtained by the channel owner through advertising?
\end{inparaenum}
Below, we develop models to address each of these questions. 
\begin{compactenum}[1)]
\item \textbf{Dynamics of viewer engagement}: 
Similar to~\cite{Ask07}, viewer engagement, in the context of live channels, can be defined as the following process: 
\begin{inparaenum}[(i)]
	\item The viewer decides to watch the live channel. 
	\item The viewer is ``engaged'' with the content of the live channel. 
	\item The viewer will watch the live content without switching to other channels. \label{veng:benifits:1}
	\item The viewer is more attentive when watching the live content.  \label{veng:benifits:2}
\end{inparaenum}
The potential benefits of~{(iii)} and~{(iv)} above is that it increases the odds of the viewers being exposed and persuaded by advertisements. 

Viewer engagement, as defined above, is an abstract concept which captures viewer attitude, behaviour and attentiveness. 
Archak et.\ al.~\cite{AMM12,AMM10} developed a Markov chain model for online behaviour of users and the effects of advertising. 
The main finding is that the user behaviour in online social media can be approximated by a first-order Markov chain. 
Following Archak et.\ al.~\cite{AMM12,AMM10}, we model the \veng at time $t$, denoted by $\Root_\Time$, as an $\statedim$-state Markov chain with state-space $\statespace \equiv \left\{1,2, \cdots, {\statedim} \right\}$. 
The dynamics of the \veng of the channel, modelled as a Markov chain, can be characterized by the transition matrix $\tp$ and initial probability vector $\belief_0$ as follows:
\begin{equation}
	\label{eq:rootdynamics}
	\tp(i,j) = \prob(\Root_{\Time+1} = j|\Root_\Time = i),  \quad  \belief_0(i) = \prob(\Root_0 = i)
\end{equation}
The Markov chain model for the \veng of the channel is validated using simulations in Sec.~\ref{sec:numerical:results}. 
\item \textbf{Dynamics of channel viewers}: 
The number of viewers at time $t$ depend on the \veng of the live channel. 
As viewers are more engaged with the content, they are less likely to switch the channel. 
Hence, a higher viewer engagement state has higher number of viewers compared to a lower engagement state. 
Therefore, we model the dynamics of channel viewers as follows: 
The number of viewers at time $t$, denoted by $\obs_\Time$, belongs to the countably infinite set $\obspace$ of non-negative integers. 
Denote, the conditional probability of $j$ viewers ($\obs_t=j$) in \veng state $i$ ($X_t=i$) by $B(i,j)$. 
Note that the conditional probability $B(i,j)$ is assumed to be time homogeneous\footnote{The conditional probability $B(i,j)$ does not depend on the time index, $t$. }. 
The number of viewers $\obs_\Time$ is modeled as a Poisson random variable with state dependent mean $\g_{i}, i \in \statespace$, based on evidence in \cite{HZZZ06} and \cite{sta87}, as follows: 
\begin{equation}
	\label{eq:obprob}
		\obp(i,j) = \prob\left(\Obs_{\Time} = j|\state_\Time = {i} \right) = \frac{g_i^j \exp{(-g_i)}}{j!}, ~ \forall i \in \statespace,j \in \obspace. 
\end{equation} 
The states with higher state dependent mean correspond to states with higher viewer engagement. 

The channel owner does not observe the true \veng of the channel, $X_t$. 
However, at each time instant $t$, the channel owner receives a noisy observation of the \veng of the channel by the number of viewers, $\obs_\Time$. 
Hence, the channel owner needs to estimate the \veng using the history of noisy observations and schedule ads. 
\item \textbf{Reward of channel owner}: 
The channel owner agrees to show $L$ ads during the live session, which are decided prior to the beginning of the session. 
For example, the ads during the Super Bowl 50 in YouTube Live had to be pre-booked in advance.\footnote{\url{http://www.campaignlive.co.uk/article/youtube-launches-real-time-ads-major-live-events-starting-super-bowl-50/1380260}}
Hence, at each time instant $t$, the channel owner chooses an action $\action_\Time$ as follows: The channel owner can continue with the live session (denoted by \emph{Continue}) or can pause the live session to insert an ad (denoted by \emph{Stop}). 
Hence, $\action_\Time \in \actionspace=\{\text{Stop(1)},\text{Continue(2)}\}$\footnote{The Stop and Continue actions will be denoted by $1$ and $2$, respectively in the remainder of the paper.} denote the actions available to the channel owner at time~$\Time$. 
This problem of scheduling the $L$ ads opportunistically, so as to obtain maximum revenue, corresponds to a multiple stopping problem with $L$ stops. 

Choosing to stop at time $\Time$ (and schedule an ad), with $l$ stops remaining ($l$ ads remaining), the channel owner will accrue a reward $r_l(\Root_t,a=1)$, where $\Root_t$ is the state of the Markov chain at time $t$. The reward obtained by the channel owner depends on two factors: \begin{inparaenum}[(i)]\item the number of viewers \item the completion rate  and click rate  of the ad\end{inparaenum}. To capture these, we model the reward as follows:   
\begin{equation}
	r_l(\Root_t=i,a=1) = \alpha_i g_i. 
	\label{eqn:reward:stop}
\end{equation}
In~\eqref{eqn:reward:stop}, $g_i$ captures the average number of viewers in any \veng state. The term $\alpha_i \in \left[0,1\right]$ captures the completion and click rate of the ads at any \veng state. 
Similarly, if the channel owner chooses to continue, he will accrue $r_l(\Root_t,a=2)$. When an ad is not shown, the reward obtained by the channel owner is usually zero, hence, $r_l(\Root_t,a=2)=0$. 

Hence, to maximize revenue, the channel owner needs to opportunistically schedule ads at time slots when the \veng is high, corresponding to a higher number of viewers and higher click rate.  

\end{compactenum}
%
%
%

\subsection{Ad Scheduling : Problem formulation \& Stochastic Dynamic Programming}
\subsubsection{Problem Formulation}
The ad scheduling policy (or the control policy) prescribes a decision rule that determines the action taken by the channel owner. 
Let the initial probability vector, $\pi_0$ and the history of past observations at time $t$ for the channel owner be denoted as
	$
		\history_\Time = \left\{\belief_0,\obs_{1}, \cdots, \obs_{\Time} \right\}
	$. 
The control policy, at time $t$, maps the history $\history_\Time$ to action. Hence, the policy of the channel owner $\policy$ belongs to the set of admissible policies  
$
\policyspace = \left\{\policy : \policy ~ \text{maps} ~ \history_\Time \rightarrow \actionspace \right\}.
$ 
Below, we reformulate the sequential multiple stopping problem of scheduling ads in terms of belief state. 

Let $\Belief = \left\{\belief \in \Real^{\statedim}: \one_{\statedim}^\prime\belief = 1, \belief(i) \geq 0  \right\}$ denote the belief space of all $S$-dimensional probability vectors. 
The belief-state $\belief_\Time$ is a sufficient statistic of $\history_\Time$~\cite{bertsekas1995dynamic}, and evolves as $\belief_{\Time+1} = \filter(\belief_{\Time},\obs_{\Time+1})$, where
\begin{align} 
	\label{eq:hmmfilter}
	\begin{aligned}
		&\filter(\belief_{\Time},\obs_{\Time+1}) = \cfrac{\obp_{\obs_{\Time+1}}\tp^{'}\belief_{\Time}}{\filternorm(\belief_{\Time},\obs_{\Time+1})}, \quad \filternorm(\belief_{\Time},\obs_{t+1}) = \one_{\statedim}^{\prime}\obp_{\obs_{\Time+1}}\tp^{'}\belief_{\Time}, \\
&\obp_{\obs_{\Time+1}} = \text{diag}\left(\obp(1,\obs_{\Time+1}), \cdots, \obp(\statedim,\obs_{\Time+1})\right).
\end{aligned}
\end{align}
Here $\one_\statedim$ represents the $\statedim$-dimensional vectors of ones. 

The aim is to compute the optimal stationary ad scheduling policy  
$	\action_t = \mu(\belief_t, l), $
as a function of the belief, $\belief_t$, and the number of stops (or the number of ads) remaining, $l$, to maximize the infinite horizon criterion defined in~\eqref{eq:discountedreward_h}. 
Let $\tau_l$ denote the stopping time when there are $l$ stops remaining:  
\begin{equation}
	\tau_l = \inf \left\{t: t > \tau_{l+1}, \action_t = 1 \right\}, \text{with } \tau_{L+1} = 0. 
	\label{eqn:stopping:time:l}
\end{equation}
The infinite horizon discounted criterion with stationary policy $\policy$, and initial belief $\belief_0$ is as below: 
\begin{align}
	\discountedreward_{{\policy}}(\belief_0) & = \E_\mu\left\{\sum_{\Time=0}^{\tau_L-1}\discount^{\Time}r_L(\state_\Time,2) + \discount^{\tau_L}r_L(\state_{\tau_L},1)+ \sum_{\Time={\tau_L+1}}^{\tau_{L-1}-1}\discount^{\Time}r_{L-1}(\state_\Time,2)+ \dots +    \discount^{\tau_1}r_1(\state_{\tau_1},1) \given[\Big] \pi_0 \right\}, \label{eq:discountedreward_h} \\
	&= \E_\mu\left\{\sum_{\Time=0}^{\tau_L-1}\discount^{\Time}r_{2,L}^\prime \belief  + \discount^{\tau_L}r_{1,L}^\prime \belief+ \sum_{\Time={\tau_L+1}}^{\tau_{L-1}-1}\discount^{\Time}r_{2,L-1}^\prime + \dots +    \discount^{\tau_1}r_{1,1}^\prime \belief \given[\Big] \pi_0 \right\}, \label{eq:discountedreward}
\end{align}
where $r_{\action, l} = \left[r_l(1,\action), \dots, r_l(\statedim,\action)\right]^\prime$. In~\eqref{eq:discountedreward_h} and~\eqref{eq:discountedreward}, $\rho \in \left(0,1\right]$ denotes the discount factor. Below, we study the special case, where $r_{1,1}=r_{1,2}=\dots=r_{1,L} = r$ and $r_{2,1}=r_{2,2}=\dots=r_{2,L} = 0$, i.e.\ the reward for stopping and scheduling an ad is independent of the number of stops remaining and the channel owner accrues no reward for continuing. The extension to the general case is straightforward.  
%
\subsubsection{Stochastic Dynamic Programming}
The computation of the optimal ad scheduling policy $\optpolicy$, to maximize the infinite horizon discounted criterion in~\eqref{eq:discountedreward_h} and~\eqref{eq:discountedreward}, is equivalent to solving Bellman's dynamic programming equation~\cite{bertsekas1995dynamic}: 
\begin{equation}
\label{eq:bellman}
\optpolicy(\belief,\looks) = \underset{\action \in \actionspace}{\argmax} ~ Q(\belief,\looks,\action), \quad
\valuefunction(\belief,\looks) = \underset{\action \in \actionspace}{\max} ~ Q(\belief,\looks,\action),  
\end{equation}
where,
\begin{equation}
		Q(\belief,\looks, 1) 	=    r^\prime\belief + \discount\sum_{\obs \in \obspace}\valuefunction\left(\filter(\belief,\obs),\looks-1\right)\filternorm(\belief,\obs), \quad
		Q(\belief,\looks, 2) 	= 	\discount\sum_{\obs \in \obspace}\valuefunction\left(\filter(\belief,\obs),\looks\right)\filternorm(\belief,\obs) 
\end{equation} 

The above dynamic programming formulation is a POMDP. Since the state-space $\Belief$, is a continuum, Bellman's equation \eqref{eq:bellman} does not translate into practical solution methodologies as $\valuefunction(\belief,\looks)$ needs to be evaluated at each $\belief \in \Belief$.  This in turn renders the calculation of the optimal policy $\optpolicy(\belief,\looks)$ computationally intractable. In Sec.~\ref{sec:scheduling:policy} we derive structural results of the optimal ad scheduling policy. The advantage of deriving structural results is that the optimal policy can be computed efficiently. Sec.~\ref{sec:spsa:MLR:threshold} provides stochastic approximation algorithms to compute approximations of the optimal policy using the structural results derived in Sec.~\ref{sec:scheduling:policy}. 
\section{Optimal ad scheduling: Structural results }
\label{sec:scheduling:policy}
In this section, we derive structural results for the optimal ad scheduling policy. 
We first introduce the value iteration algorithm in Sec.~\ref{subsec:value:iteration}, a successive approximation method to solve the dynamic programming recursion in~\eqref{eq:bellman}. 
The value iteration algorithm is a valuable tool for deriving the structural results. 
In Section~\ref{subsec:strucutral:results}, we use the value iteration algorithm in Sec.~\ref{subsec:value:iteration} to prove structural results of the optimal ad scheduling policy.  
Using the structural results in Sec.~\ref{subsec:strucutral:results}, we provide a simulation based algorithm to compute the policy in Sec.~\ref{sec:spsa:MLR:threshold}. 
\subsection{Value Iteration Algorithm}
\label{subsec:value:iteration}
The value iteration algorithm is a successive approximation approach for solving Bellman's equation~\eqref{eq:bellman}. 

The procedure is as follows: For iterations $k = 0,1,\dots$, the value function $V_k(\pi,l)$ and the policy $\mu_k(\pi,l)$ is obtained as follows
\begin{equation}
	\label{eqn:dyn}
	V_{k+1}(\pi,l) = \underset{u \in \{1,2\}}{\text{max}}  Q_{k+1}(\pi,l,u), 
\end{equation}
\begin{equation}
	\label{eqn:dyn:policy}
	\mu_{k+1}(\pi,l) = \underset{u \in \{1,2\}}{\argmax }  Q_{k+1}(\pi,l,u), 
\end{equation}
where
\begin{equation}
	Q_{k+1}(\pi,l,1) = \reward{\pi} + \rho \sum_{y} V_{k}(\Tpiy,l-1)\Spiy, 
	\label{eqn:Q:stop}
\end{equation}
and
\begin{equation}
	Q_{k+1}(\pi,l,2) = \rho \sum_{y} V_{k}(\Tpiy,l)\Spiy,  
	\label{eqn:Q:continue}
\end{equation}
with $V_{0}(\pi,l)$ initialized arbitrarily. 
It can be easily shown that the above procedure converges~\cite{bertsekas1995dynamic}. 

In order to prove the structural results of the stationary ad scheduling policy, defined in Sec.~\ref{sec:system:model}, we define the stopping and the continuance regions of the policy as below. 
Let $W_{k}(\pi,l)$ be defined as 
\begin{equation}
	W_{k}(\pi,l) \triangleq V_{k}(\pi,l)-V_{k}(\pi,l-1).
	\label{eqn:def:W:k}
\end{equation}

The stopping and continuance region (at each iteration $k$) is defined as follows:
\begin{align}
	\begin{aligned}
		{S_{k+1}^{l}} = \{ \pi | \reward{\pi} \geq \rho \sum_y W_{k}(\Tpiy,l)\Spiy \}  \\
		{C_{k+1}^{l}} = \{ \pi | \reward{\pi} < \rho \sum_y W_{k}(\Tpiy,l)\Spiy \} 
	\end{aligned}
\label{eqn:continueset}
\end{align} 
Since the value iteration converges, the optimal stationary policy $\mu^*(\pi,l)$ is defined as 
\begin{equation}
	\mu^*(\pi,l) = \lim_{k \to \infty} \mu_{k}(\pi,l).
	\label{eqn:value:iteration:limit:policy}
\end{equation}
Correspondingly, the stationary stopping and continuance sets are defined by
\begin{equation}
	S^l = \lim_{k \to \infty} S_{k}^l, \quad C^l = \lim_{k \to \infty} C_{k}^l.
	\label{eqm:value:iteration:limit:sets}
\end{equation}
The value function, $V_{k}(\pi,l)$ in~\eqref{eqn:dyn}, can be rewritten, using~\eqref{eqn:continueset}, as follows:
\begin{equation} 
	V_{k}(\pi,l) = {\left(\reward{\pi} + \rho \sum_{y} V_{k-1}(\Tpiy,l-1)\Spiy\right)\mathcal{I}_{S_{k}^{l}}} + {\left(\rho \sum_{y} V_{k-1}(\Tpiy,l)\Spiy\right)\mathcal{I}_{C_{k}^{l}}} 
	\label{eqn:V:k:pi:l}
\end{equation}
where $\mathcal{I}_{C_{k}^{l}}$ and $\mathcal{I}_{S_{k}^{l}}$ are indicator functions on the continuance and stopping regions respectively, for each iteration $k$. 

Assume $S_{k}^{l-1} \subset S_{k}^{l}$ (see Proposition~\ref{prop:nested}) and substituting~\eqref{eqn:V:k:pi:l} in the definition of $W_{k}(\pi,l)$ in~\eqref{eqn:def:W:k}, 
\begin{align}
  W_{k}(\pi,l)  &= \left(\rho \sum_{y} W_{k-1}(\Tpiy,l)\Spiy\right) \mathcal{I}_{C_{k}^{l}}(\pi) \nonumber \\
  		&+ {\reward{\pi} \mathcal{I}_{C_{k}^{l-1}\cap S_{k}^{l}}(\pi)} \label{w_k_l} \\
		&+ \left(\rho \sum_{y} W_{k-1}(\Tpiy,l-1)\Spiy\right) \mathcal{I}_{S_{k}^{l-1}}(\pi) \nonumber
\end{align}

The value iteration algorithm~\eqref{eqn:dyn} to~\eqref{eqn:Q:continue} does not translate into a practical solution methodology since the belief state belongs to an uncountable set. 
Hence, there is a strong motivation to characterize the structure of the optimal policy.  
In the following section, we use the definition of $W_{k}(\pi,l)$ in~\eqref{w_k_l} to prove structural results on the stopping region defined in~\eqref{eqn:continueset}.
\subsection{Structural results for the optimal ad scheduling policy}
\label{subsec:strucutral:results}
In this section, we provide structural results on the optimal policy of the multiple stopping problem corresponding to maximizing the ad revenue. 

\subsubsection{Assumptions}
\label{subsubsec:assumptions}
The main result below, namely, Theorem~\ref{thm:main}, requires the following assumptions on the reward vector, $r$, and the transition matrix, $P$ and the observation distribution, $B$. 
The first assumption on the reward says that $e_1$ is the state with the highest reward and the reward monotonically decreases. 
This captures the channel owners' preference of scheduling ads at the highest \veng state. 
A sufficent condition for the reward to monotonically decrease is that both the mean number of viewers $g_i$, and the completion and click rate $\alpha_i$ be non-increasing. 
The second and the third assumptions {(A2)} and {(A3)} relate to the underlying stochastic model and is related to MLR ordering of the updated belief vector in~\eqref{eq:hmmfilter} (see Theorem~\ref{thm:VK:filter} and Theorem~\ref{thm:VK:Value} in Appendix~\ref{appendix:def:mlr:lines}). The assumption~{(A2)} models the following facts: \begin{inparaenum}[i)] \item The user behaviour in online social media can be approximated by a first order Markov chain. \item The \veng changes at a smaller time scale compared to sampling or decision epochs. \end{inparaenum}
The assumption~{(A3)} is due to the fact that the \veng states can be ordered corresponding to decreasing mean number of viewers, $g_i$. 
The last assumption~{(A4)} is a technical condition required for our proof. 
\begin{itemize}
	\item(A1) The vector, $r$, has decreasing elements. Hence, $\reward{\pi}$ is increasing in $\pi$. \label{ass:u}
	\item(A2) $B$ is TP2\footnote{Refer to Appendix~\ref{appendix:def:mlr:lines} for the definition of TP2 ordering. \label{footnote:TP2:order}}. A necessary and sufficient condition for $B$ to be TP2 is that the state dependent mean $g_i$ in~\eqref{eq:obprob} is monotonically decreasing. \label{ass:observation}
	\item(A3) $P$ is TP2\footref{footnote:TP2:order}. \label{ass:transition}
	\item(A4) The vector, $(I-\rho P^\prime)r$, has decreasing elements.  
\end{itemize}

\subsubsection{Main Result}
\begin{figure}[h]
	\centering
	\scalebox{0.5}{\input{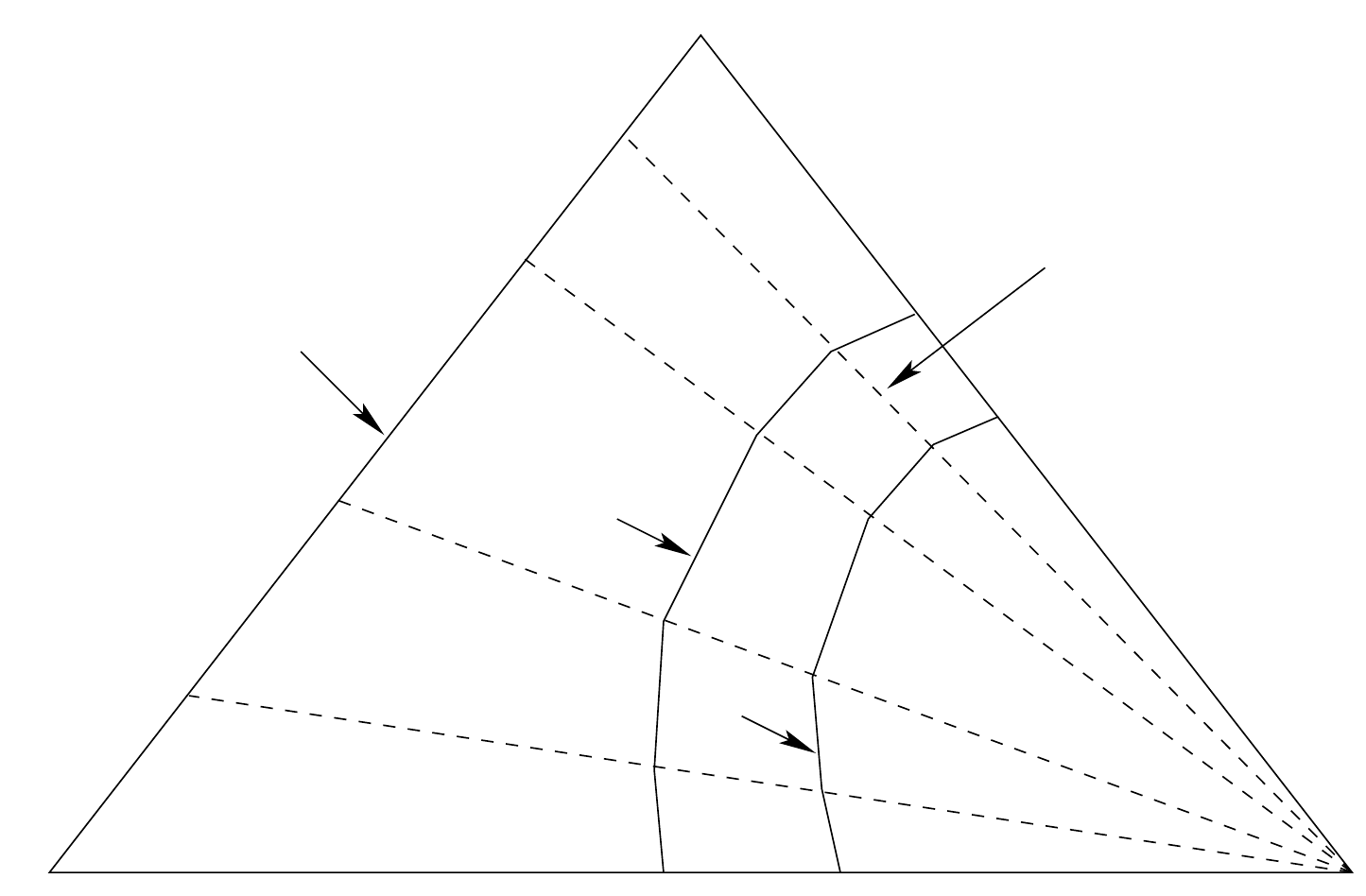}}
	\caption{$\mathcal{L}(e_1,\bar{\pi})$ corresponds to a line joining $e_1$ and any $\bar{\pi} \in \mathcal{H}$\protect\footnotemark on the simplex $\Pi$. The advantage of MLR ordering on lines is that the belief states on the line $\mathcal{L}(e_1,\bar{\pi})$ and $\mathcal{L}(e_X,\bar{\pi})$ can be fully ordered. Hence, we can define monotonicity of the policy over the lines. This is not possible for MLR ordering, since it is a partial order. }
	\label{fig:mlr:lines}
\end{figure}

The following (Theorem~\ref{thm:main}) is the main result of the paper and the proof is provided in Appendix~\ref{appendix:proofs}.  
Theorem~{\ref{thm:main}.\ref{thm:list:monotone:result}} states that the optimal policy is a monotone policy. The optimal policy $\mu^*(\pi,l)$ decreases monotonically with the belief state $\pi$. 
However, for a monotone policy to be well defined, we need to first define the ordering between two belief states. 
In this paper, we use the Monotone Likelihood Ratio (MLR) ordering\footnote{MLR ordering is defined in Def.~\ref{def:mlr:ordering} in Appendix~\ref{subsec:mlr:ordering}}, and the less restrictive MLR ordering on \emph{lines} $\mathcal{L}(e_1,\bar{\pi})$ and $\mathcal{L}(e_S,\bar{\pi})$\footnote{MLR ordering over lines is defined in Appendix~\ref{subsec:mlr:ordering:lines}} over the belief states~\cite{KD07,krishnamurthy2016partially}. Fig.~\ref{fig:mlr:lines} illustrates the definition of $\mathcal{L}(e_1,\bar{\pi})$. 
\footnotetext{$\mathcal{H}$ is defined in Appendix~\ref{subsec:mlr:ordering:lines}}
\begin{theorem}
	Assume {(A1)}, {(A2)}, {(A3)} and {(A4)}, then, 
	\begin{enumerate}[label=\Alph*]
		\item \label{thm:list:monotone:result}There exists an optimal policy $\mu^*(\pi,l)$ that is monotonically decreasing on lines $\mathcal{L}(e_1,\bar{\pi})$, and $\mathcal{L}(e_X,\bar{\pi})$ for each $l$.
		\item \label{thm:list:connected:sets}There exists an optimal switching curve $\Gamma_l$, for each $l$, that partitions the belief space $\Pi(X)$ into two individually connected regions $S^l$ and $C^l$,such that the optimal policy is 
			\begin{equation}
				\mu^*(\pi,l) = 	\begin{cases}
					1 & \text{if } \pi \in S^l \\ 
					2 & \text{if } \pi \in C^l 
				\end{cases}
				\label{eqn:policy}
			\end{equation}
		\item \label{thm:list:subset}$S^{l-1} \subset S^l $. 
	\end{enumerate}
	\label{thm:main}
\end{theorem}
Theorem~\ref{thm:main}\ref{thm:list:monotone:result} implies that the optimal action is monotonically decreasing on the line $\mathcal{L}(e_1,\bar{\pi})$, as shown in Fig.~\ref{fig:mlr:lines}. Hence, on each line $\mathcal{L}(e_1,\bar{\pi})$ there exists a threshold above (in MLR sense) which it is optimal to \emph{Stop} and below which it is optimal to \emph{Continue}. Theorem~\ref{thm:main}\ref{thm:list:connected:sets} says, for each $l$, the stopping and continuance sets are connected. Hence, there exists a threshold curve, $\Gamma_l$, as shown in Fig.~\ref{fig:mlr:lines}, obtained by joining the thresholds, from Theorem~\ref{thm:main}\ref{thm:list:monotone:result}, on each of the line $\mathcal{L}(e_1,\bar{\pi})$. Theorem~\ref{fig:mlr:lines}\ref{thm:list:subset} proves the sub-setting structure of the stopping and continuance sets, as shown in Fig.~\ref{fig:mlr:lines}. 

In order to prove the theorem, we introduce the following propositions, proofs of which are provided in the Appendix~\ref{appendix:proofs}. 
The value function of the classical (single) stopping POMDP increases with $\pi$ (in MLR sense)~\cite{krishnamurthy2016partially,Kri13,Kri11}. 
Proposition~\ref{prop:V:increase:pi} states that the above result holds even in the multiple stopping problem.  
In the classical stopping POMDPs in~\cite{krishnamurthy2016partially}, the stopping and continuance sets are characterized using the value function. 
However, in the multiple stopping problem, considered in this paper, $W$ plays the role of value function in characterizing the stopping and continuance region. 
Proposition~\ref{prop:W:decrease:l} proves the corresponding result in the multiple stopping problem. 
Proposition~\ref{prop:nested} proves the nested stopping regions at each iteration $k$ of the value iteration. 
Since the value iteration converges, Proposition~\ref{prop:nested} implies Theorem~\ref{fig:mlr:lines}\ref{thm:list:subset}. 
\begin{proposition}
	\label{prop:V:increase:pi}
	$V_{k}(\pi,l)$ is increasing in $\pi$.
\end{proposition}

\begin{proposition}
	\label{prop:W:decrease:l}
	$W_{k}(\pi,l)$ is decreasing in $l$.
\end{proposition}

\begin{proposition}
	\label{prop:nested}
	$S_{k+1}^l \supset S_{k+1}^{l-1}$
\end{proposition}
To summarize, we showed the following properties of the optimal policy: 
\begin{compactenum}[(i)]
\item the optimal policy is monotone on the lines $\mathcal{L}(e_1,\pi)$. 
\item existence of a unique threshold curve for the stopping region, $\Gamma^l$. 
\item the stopping regions have a sub-setting property, i.e.\ $S^{l-1} \subset S^{l}$.
\end{compactenum}

\section{Stochastic Approximation Algorithm for computing best approximate ad scheduling policy}
\label{sec:spsa:MLR:threshold}
In this section, we synthesize policies satisfying the properties of the optimal ad scheduling policy derived in Sec.~\ref{sec:scheduling:policy}. 
The policies can be characterized by $L$ threshold curve, corresponding to each of the stopping regions. 
In this section, we parametrize the threshold curve, $\Gamma^l$, as $\hat{\Gamma}^l_\theta$. 
Here, $\theta$ denotes the parameter of the threshold curve. 
To capture the essence of Theorem~\ref{thm:main}, we require that the parametrized optimal policy, $\mu_\theta(l,\pi)$, be decreasing on lines $\mathcal{L}(e_1,\bar{\pi})$ and $\mathcal{L}(e_S,\bar{\pi})$ and the stopping sets are connected and satisfy the sub-setting property, i.e.\ $S^{l-1} \subset S^{l}$. Below, we will restrict our attention to obtain the best \emph{linear} threshold policy, i.e.\ policy of the form given in~\eqref{eqn:threshold:policy}. We characterize the parameters of the threshold policy in Sec.~\ref{subsec:structure:MLR:threshold} and provide an algorithm to compute the parameters using stochastic approximation in Sec.~\ref{subsec:spsa:linear:threshold:policy}. 
\subsection{Structure of best linear MLR threshold policy for ad scheduling}
\label{subsec:structure:MLR:threshold}
Consider a threshold hyperplane, on the simplex $\Pi$, of the form~\eqref{eqn:threshold:policy} where $\theta_l \in \reals^{S-1}$ denotes the coefficient vector. The linear threshold scheduling policy, denoted by $\mu_{\theta}(l,\pi)$ is defined as 
\begin{equation}
	\mu_\theta(l,\pi) = 	
	\begin{cases}
		1 & \text{if } \begin{bmatrix} 0 & 1 & \theta_l \end{bmatrix} \begin{bmatrix} \pi \\ -1\end{bmatrix} \le 0 \\
		2 & \text{else },
	\end{cases}
	\label{eqn:threshold:policy}
\end{equation}
Here, $\theta = \left(\theta_1,\theta_2,\dots,\theta_L\right) \in \reals^{L \times (S-1)}$ is the concatenation of the $\theta_l$ vectors.  

To capture the essence of Theorem~\ref{thm:main}\ref{thm:list:monotone:result}, we require that the policy be decreasing on lines, i.e.\ for $\pi_1 \ge_{\mathcal{L}_1} \pi_2,\; \mu_{\theta}(\pi_1,l) \le \mu_{\theta}(\pi_2,l)$. 
Theorem~\ref{thm:coefficent:MLR:condition} gives necessary and sufficient conditions on the coefficient vector $\theta_l$ such that the above condition holds. 
\begin{theorem}
	A necessary and sufficient condition for the linear threshold policy $\mu_\theta(l,\pi)$ to be 
	\begin{enumerate}
		\item MLR decreasing on line $\mathcal{L}(e_1,\bar{\pi})$, iff $\theta_l(S-1) \ge 0$ and $\theta_l(i) \ge 0, \; i \le S-2$. 
		\item MLR decreasing on line $\mathcal{L}(e_S,\bar{\pi})$, iff $\theta_l(S-1) \ge 0$, $\theta_l(S-2) \ge 1$ and $\theta_l(i) \le \theta_l(S-2), \; i < S-2$. 
	\end{enumerate}
	\label{thm:coefficent:MLR:condition}
\end{theorem}
The proof of Theorem~\ref{thm:coefficent:MLR:condition} is similar to the proof of Theorem~{12.4.1} in~\cite{krishnamurthy2016partially} and hence is omitted in this paper. 


The stopping sets are connected since we parametrize the threshold curve using a linear hyperplane. Finally, the linear threshold approximation curve needs to satisfy the sub-setting property in Theorem~\ref{fig:mlr:lines}\ref{thm:list:subset}. 
Theorem~\ref{thm:constraints:parameter} provides sufficient conditions such that the parametrized linear threshold curve satisfy the sub-setting property and the proof is provided in the Appendix~\ref{appendix:proof:thm:constraints:parameter}.  
\begin{theorem}
	To satisfy the sub-setting structure in Theorem~\ref{thm:main}\ref{thm:list:subset}, the parameters of the linear threshold curve have to satisfy the following condition
	\begin{equation}
		\begin{aligned}
			\theta_{l-1}(S-1) &= \theta_l(S-1) \\
			\theta_{l-1}(i) &\ge \theta_l(i) \quad i < S-1
		\end{aligned}
		\label{eqn:cond:parameter}
	\end{equation}
	\label{thm:constraints:parameter}
\end{theorem}
Therefore, under the conditions of Theorem~\ref{thm:coefficent:MLR:condition} and Theorem~\ref{thm:constraints:parameter} the linear threshold policy in~\eqref{eqn:threshold:policy} satisfy all the conditions in Theorem~\ref{thm:main} and hence qualify as the \emph{best} linear threshold policy. 

The parameter $\theta$ can be re-parametrized as follows:
\begin{equation}
	\theta^\phi_l(i) = 
	\begin{cases}
		\phi_1^2(S-1) & i = S-1 \\
		1 + \phi_1^2(S-2) \prod_{\ell=2}^l \sin^2(\phi_\ell(S-2)) & i=S-2 \\
		\left(1 + \phi_1^2(S-2) \prod_{\ell=2}^l \sin^2(\phi_\ell(S-2))\right) \sin^2(\phi_l(i)) & i < S-2 
	\end{cases}
	\label{eqn:reparameterized:theta:phi}
\end{equation}
It can be easily checked the parametrization in~\eqref{eqn:reparameterized:theta:phi} satisfies the conditions in Theorem~\ref{thm:coefficent:MLR:condition} and Theorem~\ref{thm:constraints:parameter}. 

Theorem~\ref{thm:coefficent:MLR:condition} and Theorem~\ref{thm:constraints:parameter} characterize the parameters of the linear threshold policy. 
In Sec.~\ref{subsec:spsa:linear:threshold:policy} we provide an algorithm to compute the best linear threshold policy satisfying Theorem~\ref{thm:coefficent:MLR:condition} and Theorem~\ref{thm:constraints:parameter}.  
\subsection{Simulation-based stochastic approximation algorithm for estimating the best linear MLR threshold policy for ad scheduling}
\label{subsec:spsa:linear:threshold:policy}
In this section, we provide an algorithm to compute the best linear thresholds satisfying the conditions in Theorem~\ref{thm:coefficent:MLR:condition} and Theorem~\ref{thm:constraints:parameter}. 
Recall, the optimal policy minimizes the average discounted reward in~\cref{eq:discountedreward_h,eq:discountedreward}. 
The optimal linear thresholds can be obtained by a policy gradient algorithm~\cite{bertsekas1995dynamic}. 
Algorithm~\ref{algo:policy:gradient:algorithm} is a policy gradient algorithm to compute the best linear threshold policy. 
In this algorithm, we approximate $\discountedreward_\policy$ in~\cref{eq:discountedreward_h,eq:discountedreward} by the finite time approximation 
\begin{equation}
	\discountedreward_N(\theta) = \E\left\{\sum_{l=1}^{L}\discount^{\tau_l}r(X_{\tau_l},1) \given[\Big] \tau_l \le N; \forall l\right\}. 
	\label{eqn:finite:time:approximation}
\end{equation}
Here, we have made explicit the dependence of the parameter vector, $\theta$, on the discounted reward and with an abuse of notation, have suppressed the dependence on the policy $\mu$. 
It can be shown that $J_N(\theta)$ is an asymptotically biased estimate of $\discountedreward_\policy$ and can be obtained by simulation for large $N$. 
\begin{algorithm}
\caption{Threshold-Based Policy Gradient Algorithm for Optimal Multiple Stopping}
\label{algo:policy:gradient:algorithm}
\begin{algorithmic}[1]
	\REQUIRE Assume the parameters of the optimal multiple stopping problem satisfy the assumptions in Theorem~\ref{thm:main}.
	\STATE Initialize: Choose initial parameters $\hat{\phi}_0$ and initial linear threshold policy $\mu_{\hat{\theta}_0}$ using~\eqref{eqn:threshold:policy}. 
	\FOR{each iterations $n=0,1,2,\dots$: }
	\STATE Evaluate cost $J_N(\theta^{\hat{\phi}_n})$ using~\eqref{eqn:finite:time:approximation} and gradient estimate $\nabla_\phi J_N(\theta^{\hat{\phi}_n})$ with policy $\mu_{\theta^{\hat{\phi}_n}}$ using~\eqref{eqn:spsa:gradient:estimate}. 
	\STATE Update the parameter vector ${{\hat{\phi}_n}}$ to ${{\hat{\phi}_{n+1}}}$ using~\eqref{eqn:spsa:inequality:parameter:update}. 
	\ENDFOR
\end{algorithmic}
\end{algorithm}

The policy gradient algorithm in Algorithm~\ref{algo:policy:gradient:algorithm} requires the gradient $\nabla_\phi J_N(\cdot)$ at each iteration. Computation of the gradient is quite difficult due to the non-linear dependence of the parameter $\phi$ on the cost function. Hence, in this paper, we estimate the gradient through a stochastic approximation algorithm. 

There are several stochastic approximation algorithms available in the literature: infinitesimal perturbation analysis\cite{10.2307/2583338}, weak derivatives~\cite{pflug2012optimization} and the SPSA algorithm~\cite{spall2005introduction}. 
In this paper, we use the SPSA algorithm and because of the constraints in Theorem~\ref{thm:constraints:parameter} we use a variant of SPSA that can handle linear inequality constraints~\cite{wang2008stochastic}. 

Following~\cite{spall2005introduction} and~\cite{wang2008stochastic}, the gradient estimate is obtained by picking a random direction $\omega_n$, at each iteration $n$. The estimate of the gradient is then given by
\begin{align}
	\hat{\nabla}_\phi J_N(\theta^{\hat{\phi}_n}) &= \frac{J_N(\theta^{\hat{\phi}_n+c_n \omega_n}) - J_N(\theta^{\hat{\phi}_n-c_n \omega_n})}{2c_n} \omega_n,\label{eqn:spsa:gradient:estimate}
	\shortintertext{where,}
	\omega_n(i) &= 	\begin{cases} 
				-1 & \text{with probability } 0.5 \\
				+1 & \text{with probability } 0.5. 
			\end{cases} \label{eqn:spsa:omega:def}
\end{align}

The equations for the parameter update are as follows~\cite{wang2008stochastic}:
\begin{equation}
	\phi_{n+1} = \phi_n - a_n \hat{\nabla}_\phi J_N(\theta^{\hat{\phi}_n}). 
	\label{eqn:spsa:inequality:parameter:update}
\end{equation}
The parameters $a_n$ and $c_n$ and $r_n$ are chosen as in~\cite{wang2008stochastic} as follows:
\begin{align}
	\begin{aligned}
		a_n &= \upvarepsilon(n+1+\varsigma)^{-\kappa} &0.5<\kappa\le 1, \quad \text{and } \upvarepsilon,\varsigma>0\\
		c_n &= \upmu(n+1)^{-\Upsilon}  &\upmu > 0
	\end{aligned}
	\label{spsa:parameters:an:cn:rn}
\end{align}
The stochastic approximation in Algorithm~\ref{algo:policy:gradient:algorithm} converges to a local minimum. Hence, it is necessary to try several initial conditions and pick the best threshold.  

\section{Numerical Results: Synthetic and Real Dataset}
\label{sec:numerical:results}
In this section, we present numerical results on synthetic and real datasets. 
First, we illustrate our analysis of Theorem~\ref{thm:main}, using synthetic data. 
Second, we demonstrate how the optimal multiple stopping framework, used for ad scheduling in live media, can be used to detect changes in ground truth using real data from online search. 
Online search is linked to advertising in television and online social media~\cite{JWCZ13}. 
Third, we show, through simulations, that the scheduling policy obtained from Algorithm~\ref{algo:policy:gradient:algorithm} outperforms conventional technique of scheduling ads in live social media.  
In this paper, we compare the scheduling policy obtained from Algorithm~\ref{algo:policy:gradient:algorithm} with two schemes: ``Periodic'' and ``Random''. 
The periodic scheme models the most common method of advertisement scheduling in pre-recorded videos in platforms like YouTube. 
In the context of live channels, Twitch uses a periodic scheduling where an ad is inserted periodically into the live channel~\cite{SOW13}. 
In contrast, in the random scheduling scheme, the ad is inserted randomly into the live channel. 
The random scheduling scheme is used as a benchmark to compare the revenue obtained through the periodic scheme and the policy obtained through Algorithm~\ref{algo:policy:gradient:algorithm}. 
\subsection{Synthetic Data}
\label{subsec:results:synthetic}
In this section, we do a simulation study based on synthetic data for the optimal multiple stopping problem. The objective is to illustrate the analysis in Theorem~\ref{thm:main}. In addition, we obtain the policy by solving the dynamic programming equations~\eqref{eq:bellman} and show through simulations that it outperforms conventional method of scheduling ads periodically. 

In this ``toy'' example, the \veng of the live channel can be categorized into $3$ states: ``Popular'', ``Interesting'' and ``Boring'', denoted by $1$, $2$ and $3$, respectively.  
The transition between the various states follow a Markov chain with transition matrix given in~\eqref{eqn:transition:matrix:sim:study}. 
The \veng state is observed through a state dependent Poisson process with mean given in~\eqref{eqn:observation:matrix:sim:study}. The reward structure for scheduling the ads is as in~\eqref{eqn:reward:sim:study}.  

\begin{minipage}{.5\linewidth}
\begin{equation}
	P = 
	\begin{bmatrix}
		0.2 & 0.1 & 0.7\\
		0.1 & 0.1 & 0.8\\
		0 & 0.1 & 0.9
	\end{bmatrix}
	\label{eqn:transition:matrix:sim:study}
\end{equation}
\end{minipage}%
\begin{minipage}{.5\linewidth}
\begin{align}
	g &= 
	\begin{bmatrix}
		12 & 7  & 2
	\end{bmatrix}
	\label{eqn:observation:matrix:sim:study} 
	\\
	r &= 
	\begin{bmatrix}
		9& 3& 1
	\end{bmatrix}
	\label{eqn:reward:sim:study}
\end{align}
\end{minipage}

\begin{figure}[h]
	\centering
	\includegraphics[width=0.75\textwidth]{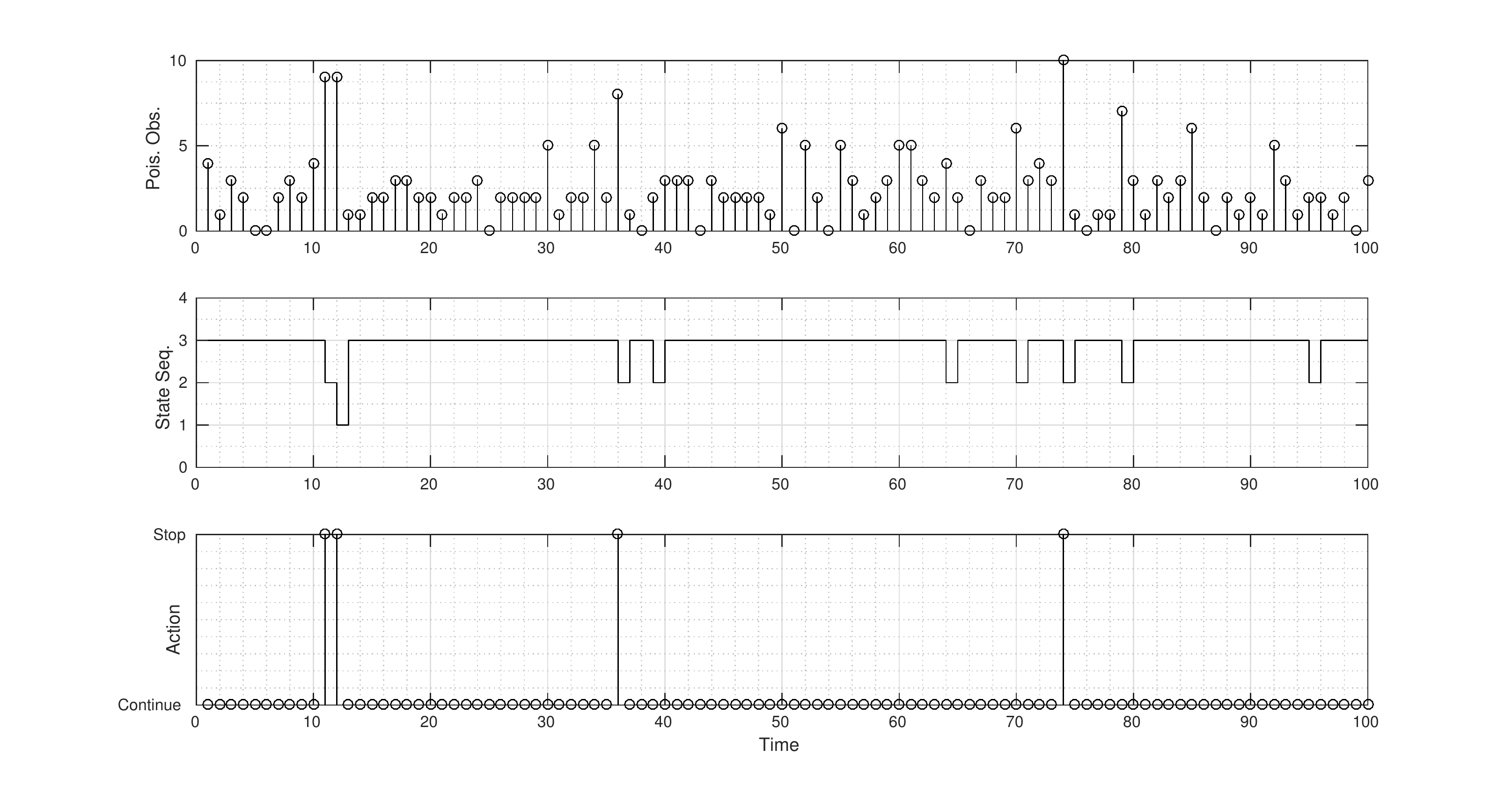}
	\caption{Sample trajectory of the state, observation sequence and the policy.}
	\label{fig:simulation:study:sample:trajectory}
\end{figure}

\begin{figure}[!ht]
  \centering
  \subfigure[Stopping set for $l=5$]{\label{fig:simulation:study:stopping:set:5} \includegraphics[scale=0.5]{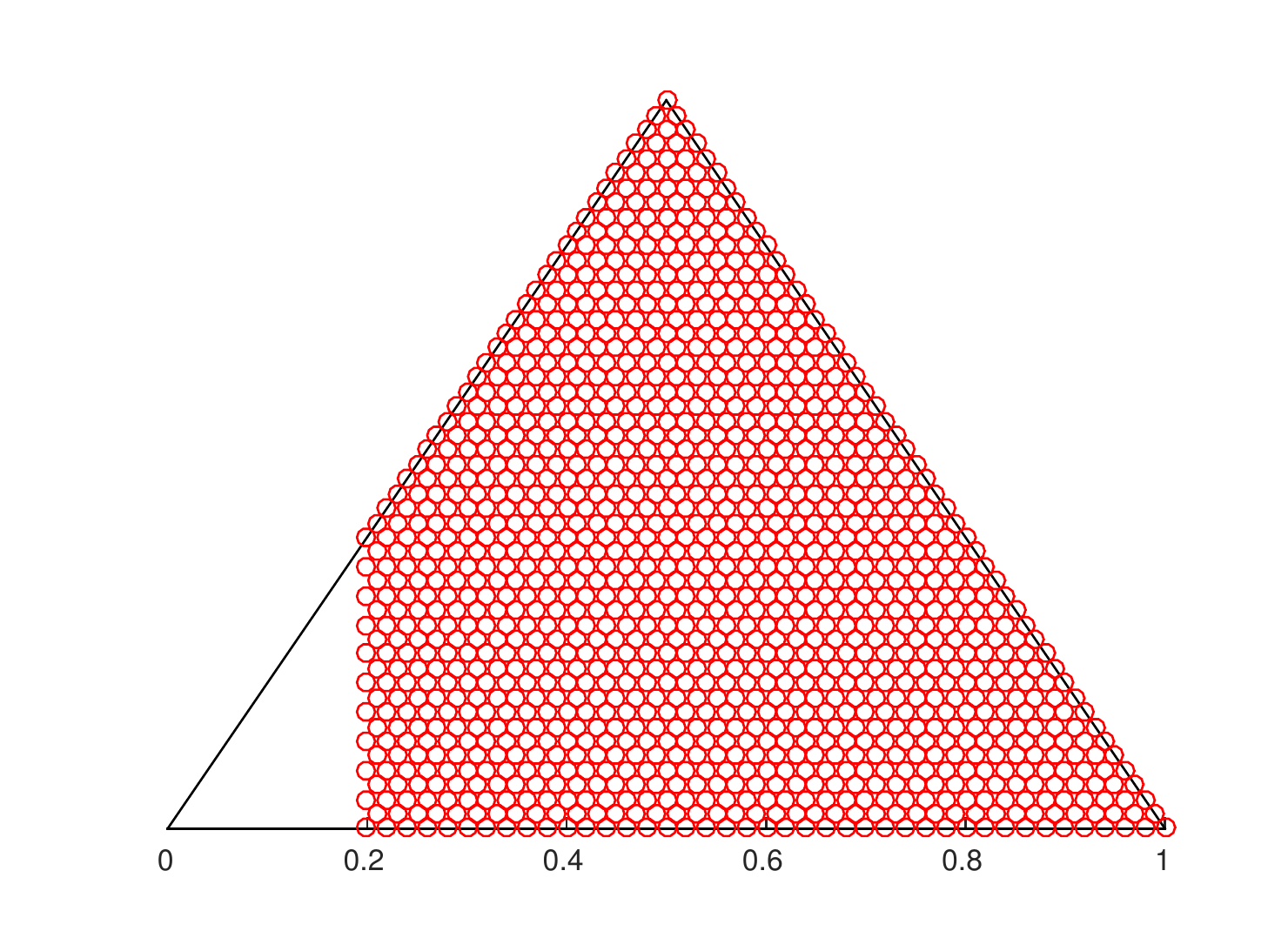}}
  \subfigure[Stopping set for $l=1$]{\label{fig:simulation:study:stopping:set:1} \includegraphics[scale=0.5]{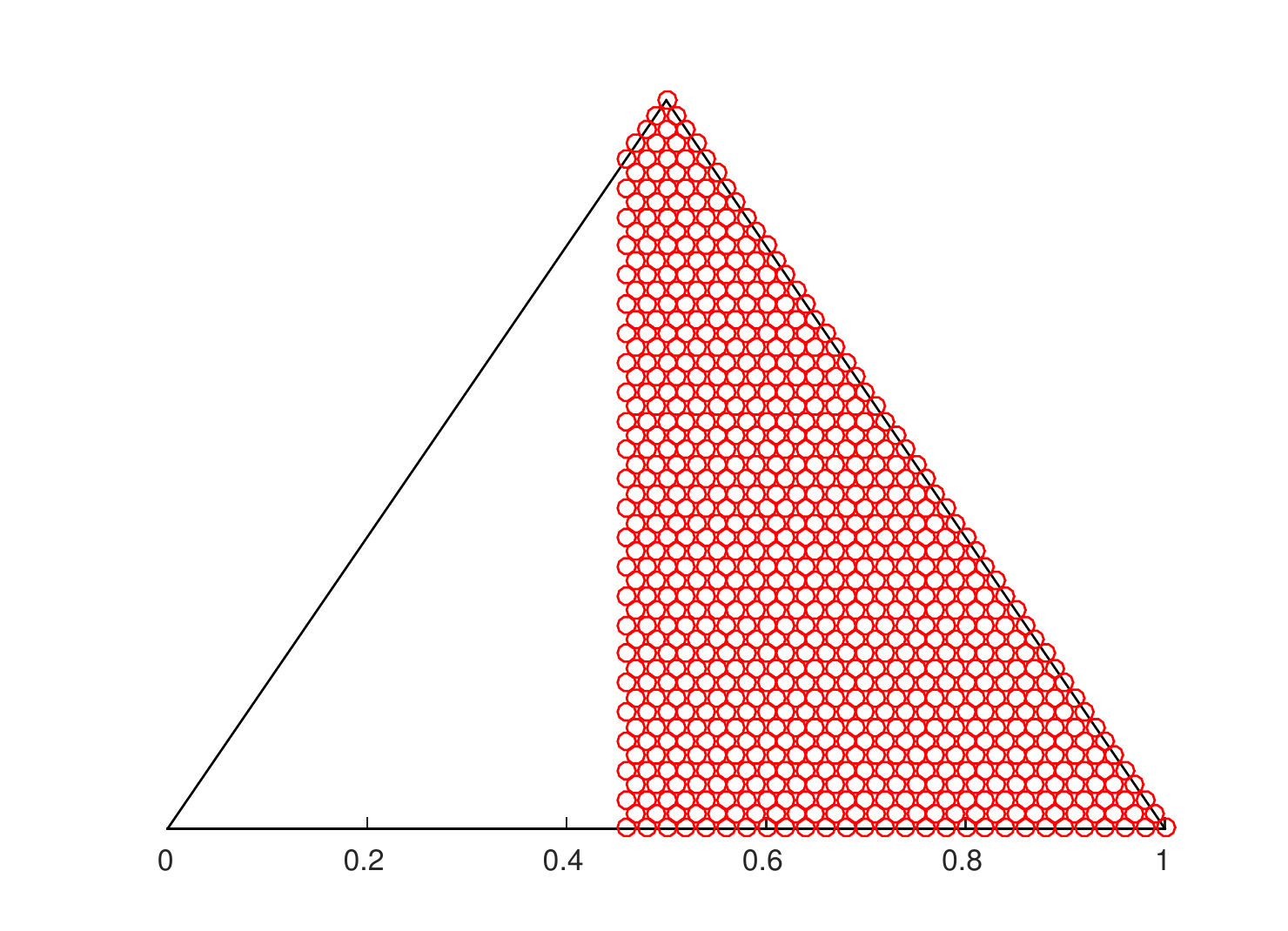}}
  \caption[Stopping Set]{Stopping set (shown in red) obtained by solving the dynamic programming in~\eqref{eq:bellman}. The figure illustrates the sub-setting structure of the stopping sets, i.e.\ $S^{l-1} \subset S^l$, in Theorem~\ref{thm:main}. }
\label{fig:simulation:study:stopping:set} 
\end{figure}

Fig.~\ref{fig:simulation:study:sample:trajectory} shows a sample trajectory of the observation and the state sequence. For $L=5$, i.e.\ for a total of $5$ ads to be scheduled, we obtained the policy by solving the dynamic programming equations in~\eqref{eq:bellman}. The resulting policy, in terms of stopping and continuance set as defined in~\eqref{eqn:def:W:k}, is shown in Fig.~\ref{fig:simulation:study:stopping:set}. 
We also show the corresponding points where the policy was chosen in Fig.~\ref{fig:simulation:study:sample:trajectory}. 
Only the stopping set for $l=5$ and $l=1$, i.e.\ $S^5$ and $S^1$ respectively are shown in Figure~\ref{fig:simulation:study:stopping:set}. 
As can be seen from Figure~\ref{fig:simulation:study:stopping:set}, $S^1 \subset S^5$, verifying Theorem~{\ref{thm:main}.\ref{thm:list:subset}}. 
The stopping regions $S^1$ and $S^5$ are connected and the optimal policy is threshold on any line $\mathcal{L}(e_1,\pi)$, verifying Theorem~{\ref{thm:main}.\ref{thm:list:connected:sets}} and Theorem~{\ref{thm:main}.\ref{thm:list:monotone:result}} respectively. 

\begin{figure}[h!]
	\centering
	\includegraphics[scale=0.5]{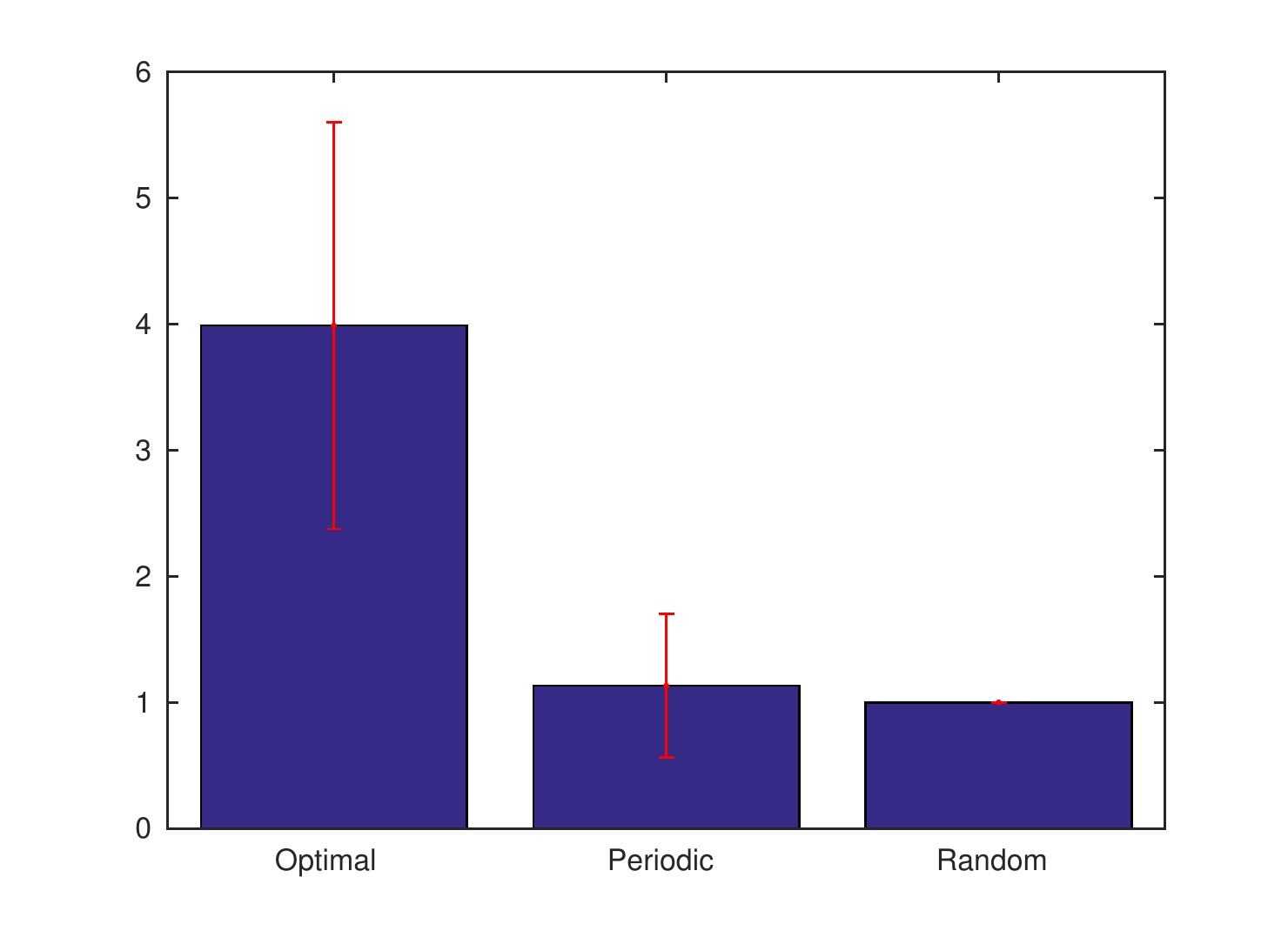}
	\caption{Comparison between the various scheduling policies: The performance of the policy obtained by solving the dynamic programming equation in~\eqref{eq:bellman} is shown by `Optimal'. The optimal policy outperforms the conventional periodic scheduling (shown as `Periodic'). The random scheduling (shown as `Random') is used for benchmarking the various scheduling policies. }
	\label{fig:simulation:study:compare:schemes}
\end{figure}

We now compare the policy obtained by solving the dynamic programming equation in~\eqref{eq:bellman} with the conventional technique of scheduling ads periodically within the live session. We also include a random scheduling policy, where the ads are scheduled randomly within the live session, for benchmarking purposes. Fig.~\ref{fig:simulation:study:compare:schemes} shows the comparison between the various schemes. The results in Fig.~\ref{fig:simulation:study:compare:schemes} was obtained by $10^4$ independent Monte Carlo (M.C.) simulations. It can be seen from Fig.~\ref{fig:simulation:study:compare:schemes} that the policy obtained from~\eqref{eq:bellman} significantly outperforms (close to $4$ times) periodic scheduling. This is not surprising since the policy obtained by solving the dynamic programming in~\eqref{eq:bellman} opportunistically schedules ads when the \veng of the channel is high. 
\subsection{Change Detection (Real Dataset)}
\label{subsec:real:data:techbuzz}
In this section, we illustrate Theorem~\ref{thm:main} for change detection problems. 
Change detection problems are special case of multiple stopping problems, when $L = 1$. 
We apply Theorem~\ref{thm:main} for detecting changes in ground truth using an online search dataset. 
Online search is linked to advertising in television and online social media~\cite{JWCZ13}. 
In addition, detecting changes in ground truth using online search data have been used for detection of outbreak of illness, political election, or major sporting events~\cite{GMPBSMB09}. 
Hence, detection of changes in ground truth is important for optimizing advertising strategy. 
The dataset that we use is the Tech Buzz dataset from Yahoo!.  
We first describe the Tech Buzz dataset in Sec.~\ref{sec:dataset:change:detection} and then show through simulations that the policy obtained by solving the dynamic programming equation in~\eqref{eq:bellman} can be used for detecting changes in ground truth using data from online search. 
\subsubsection{Dataset}
\label{sec:dataset:change:detection}
The dataset that we use in our study is the Yahoo! Buzz Game Transactions from the Webscope datasets\footnote{Yahoo! Webscope dataset: A2 - Yahoo! Buzz Game Transactions with Buzz Scores, version 1.0 \url{http://research.yahoo.com/Academic_Relations}} available from Yahoo! Labs. 
In 2005, Yahoo!~along with O'Reilly Media started a fantasy market where the trending technologies at that point where pitted against each other. 
For example, in the browser market there were ``Internet Explorer'', ``Firefox'', ``Opera'', ``Mozilla'', ``Camino'', ``Konqueror'', and ``Safari''. 
The players in the game have access to the ``buzz'', which is the online search index, measured by the number of people searching on the Yahoo!\ search engine for the technology. 
The objective of the game is to use the buzz and trade stocks accordingly.  
\subsubsection{Change Detection}
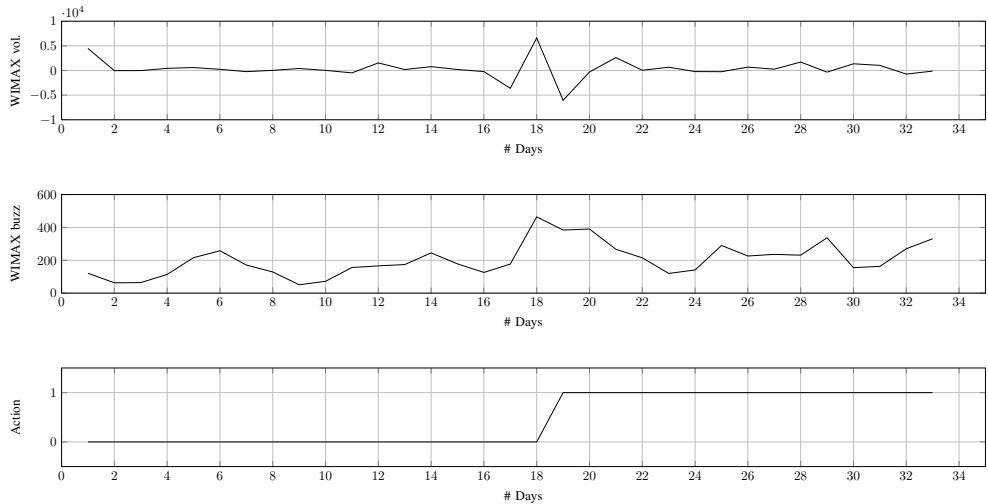
\begin{figure}[h!]
	\centering
	\scalebox{0.5}{
%
%
\begin{tikzpicture}

\begin{axis}[%
width=9.671in,
height=1.033in,
at={(1.622in,4.574in)},
scale only axis,
xmin=0,
xmax=35,
xlabel={\# Days},
xmajorgrids,
xminorgrids,
ymin=-10000,
ymax=10000,
ylabel={WIMAX vol.},
ymajorgrids,
yminorgrids,
axis background/.style={fill=white}
]
\addplot [color=black,solid,forget plot]
  table[row sep=crcr]{%
1	4471\\
2	-39\\
3	-19\\
4	439\\
5	594\\
6	236\\
7	-232\\
8	23\\
9	408\\
10	25\\
11	-484\\
12	1537\\
13	196\\
14	774\\
15	203\\
16	-219\\
17	-3611\\
18	6603\\
19	-6049\\
20	-301\\
21	2615\\
22	44\\
23	666\\
24	-220\\
25	-244\\
26	684\\
27	266\\
28	1712\\
29	-327\\
30	1358\\
31	1039\\
32	-728\\
33	-120\\
};
\end{axis}

\begin{axis}[%
width=9.671in,
height=1.033in,
at={(1.622in,2.758in)},
scale only axis,
xmin=0,
xmax=35,
xlabel={\# Days},
xmajorgrids,
xminorgrids,
ymin=0,
ymax=600,
ylabel={WIMAX buzz},
ymajorgrids,
yminorgrids,
axis background/.style={fill=white}
]
\addplot [color=black,solid,forget plot]
  table[row sep=crcr]{%
1	121\\
2	63\\
3	64\\
4	114\\
5	216\\
6	258\\
7	171\\
8	129\\
9	51\\
10	72\\
11	156\\
12	166\\
13	174\\
14	245\\
15	178\\
16	126\\
17	177\\
18	464\\
19	384\\
20	390\\
21	267\\
22	215\\
23	120\\
24	141\\
25	290\\
26	226\\
27	236\\
28	231\\
29	337\\
30	155\\
31	163\\
32	270\\
33	331\\
};
\end{axis}

\begin{axis}[%
width=9.671in,
height=1.033in,
at={(1.622in,0.941in)},
scale only axis,
xmin=0,
xmax=35,
xlabel={\# Days},
xmajorgrids,
ymin=-0.5,
ymax=1.5,
ytick={0, 1},
ylabel={Action},
ymajorgrids,
axis background/.style={fill=white}
]
\addplot [color=black,solid,forget plot]
  table[row sep=crcr]{%
1	0\\
2	0\\
3	0\\
4	0\\
5	0\\
6	0\\
7	0\\
8	0\\
9	0\\
10	0\\
11	0\\
12	0\\
13	0\\
14	0\\
15	0\\
16	0\\
17	0\\
18	0\\
19	1\\
20	1\\
21	1\\
22	1\\
23	1\\
24	1\\
25	1\\
26	1\\
27	1\\
28	1\\
29	1\\
30	1\\
31	1\\
32	1\\
33	1\\
};
\end{axis}
\end{tikzpicture}%
}
	\caption{The buzz scores and the trade volume for the WIMAX stock. The policy obtained through solving the dynamic programming in~\eqref{eq:bellman} shows a high valuation during April 18. This corresponds to Intel announcement of the WIMAX chip. The increase in valuation can also be seen by a corresponding spike in the WIMAX buzz scores. }
	\label{fig:techbuzz}
\end{figure}

We consider a subset of the data containing the WIMAX buzz scores and the number of stocks traded (volume of the stocks). 
The unknown valuation of the WIMAX technology is modelled using a $2-$state Markov chain (``1'' for high valuation and ``2'' for low valuation). 
The valuation of the stock is not observed directly, but through noisy observations on the volume of the stocks traded. 
Fig.~\ref{fig:techbuzz} shows the volume of the stocks traded and the buzz during the month of April. 
The volume of stocks traded depend on the unknown valuation and, for ease of analysis, is quantized into $3$ states (``High'', ``Medium'' and ``Low''), denoted by $1$, $2$ and $3$ respectively. 
Given the time series of the (quantized) volume of stocks traded, we obtain the hidden Markov model constituting of the transition matrix of the WIMAX valuation and the observation probability of the volume of the stocks traded given the WIMAX valuation using an EM algorithm~\cite{krishnamurthy2016partially}.  
The parameters of the Markov chain obtained using an EM algorithm are as below:

\begin{minipage}{.5\linewidth}
\begin{equation}
	P = 
	\begin{bmatrix}
	1 & 0\\
	0.1462 & 0.8538
	\end{bmatrix}
\end{equation}
\end{minipage}%
\begin{minipage}{.5\linewidth}
\begin{equation}
	B = 
	\begin{bmatrix}
		0.1489 & 0.4467 & 0.4044\\
		0.3727 & 0.5325 & 0.0947
	\end{bmatrix}
\end{equation}
\begin{equation}
	\label{eqn:change:detection:reward}
	r = 
	\begin{bmatrix}
		10 & 1
	\end{bmatrix}
\end{equation}
\end{minipage}

As can be seen from Fig.~\ref{fig:techbuzz} the WIMAX buzz and the volume of stocks traded is initially low. Hence, the objective is to find the time point at which the WIMAX value switches to the high value. 
The reward structure in~\eqref{eqn:change:detection:reward} reflects the fact that choosing to \emph{Stop} at the high value state, an agent obtains more money by trading the high value WIMAX stocks. 
The policy obtained by solving the dynamic programming in~\eqref{eq:bellman} shows a high valuation at April 18. 
The change point corresponds to Intel's announcement of WIMAX chip\footnote{\url{http://www.dailywireless.org/2005/04/17/intel-shipping-wimax-silicon/}}. 
The high valuation of WIMAX stock can also be noticed from the ``spike'' in the buzz around April 18 in Fig.~\ref{fig:techbuzz}. 
\subsection{Ad Scheduling on Live Channels (YouTube and Twitch Datasets)}
\label{subsec:real:data:youtube:twitch}
In this section, we illustrate the policy from Algorithm~\ref{algo:policy:gradient:algorithm} on real data from YouTube and Twitch. 
In comparison to Sec.~\ref{subsec:results:synthetic} and Sec.~\ref{subsec:real:data:techbuzz}, real data from YouTube and Twitch has a wide range of \veng states and hence requires more states in the Markov chain model. 
As the number of states increases, solving the dynamic programming equation in~\eqref{eq:bellman} becomes impractical. 
Hence, we resort to best linear threshold policy through Algorithm~\ref{algo:policy:gradient:algorithm}. 
We first describe the dataset in Sec.~\ref{subsec:dataset:description} and then show that the policy obtained from Algorithm~\ref{algo:policy:gradient:algorithm} outperforms conventional periodic scheduling. 
\subsubsection{Dataset}
\label{subsec:dataset:description}
In this paper, we use the dataset in~\cite{Pires:2015}\footnote{The dataset is available from \url{http://dash.ipv6.enstb.fr/dataset/live-sessions/}}.  
The dataset consists of live session on the two popular live broadcasting platforms: YouTube Live and Twitch, between January and April 2014. 
The dataset contains samples of the live sessions sampled at a $5$-minute interval on each of the platforms. 
Each sample contains the identification of the channel, the number of viewers and some additional meta-data of the channel. 
The main finding in~\cite{Pires:2015} is that the \veng is more heterogeneous than in other user-generated content platforms such as YouTube. 
The heterogeneity of \veng in live channel can be used to opportunistically schedule advertisements.  
\subsubsection{Entertainment (YouTube Live)}
\label{sec:dataset:youtube:live}
In this section we use real data from YouTube Live channel and show that the policy obtained from Algorithm~\ref{algo:policy:gradient:algorithm} outperforms conventional periodic scheduling.  

\begin{figure}[!ht]
  \centering
  \subfigure[Plot of the viewers for one session]{\label{fig:plot:viewers:category:1} \includegraphics[scale=0.45]{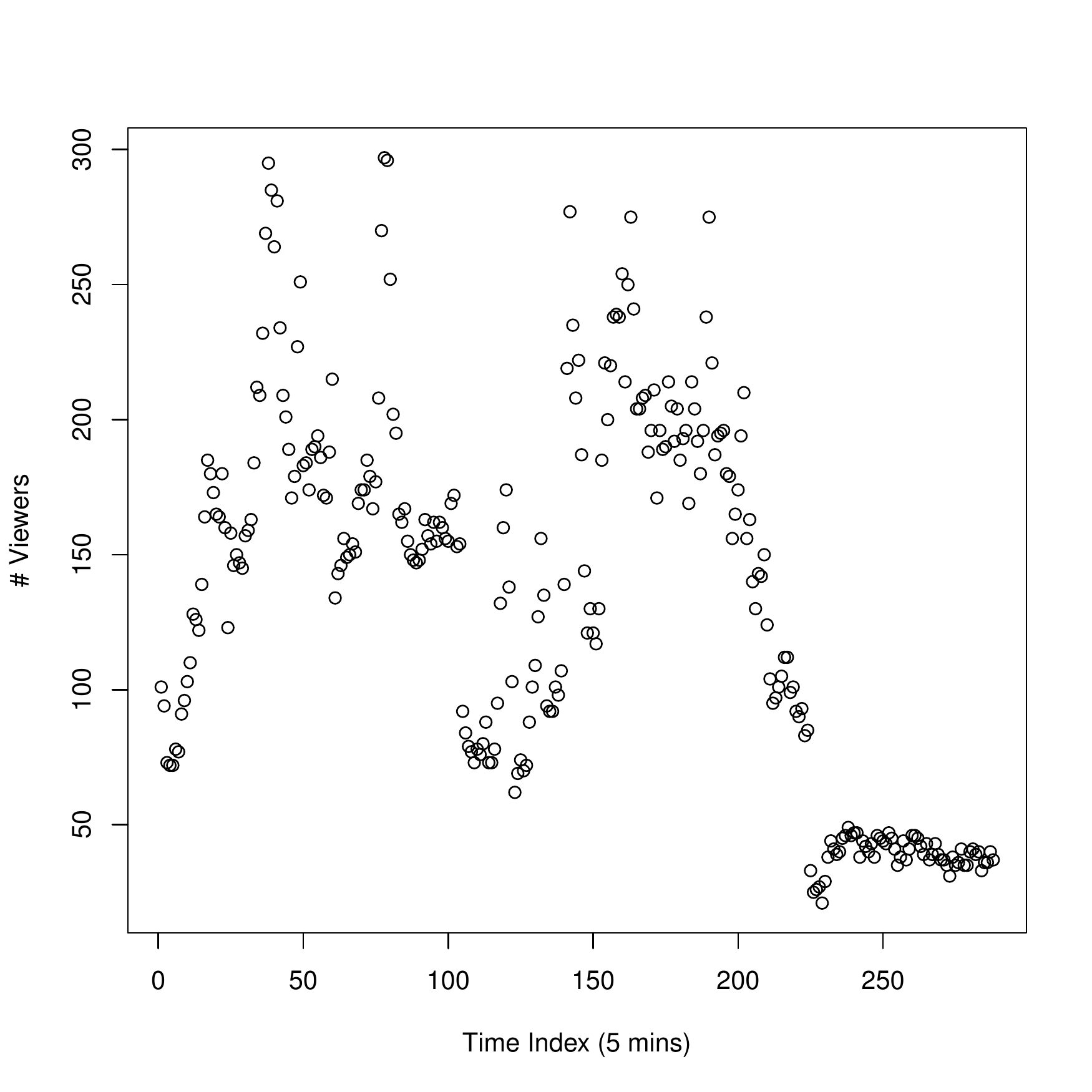}}
  \subfigure[QQ-plot]{\label{fig:qqplot:youtube} \includegraphics[scale=0.45]{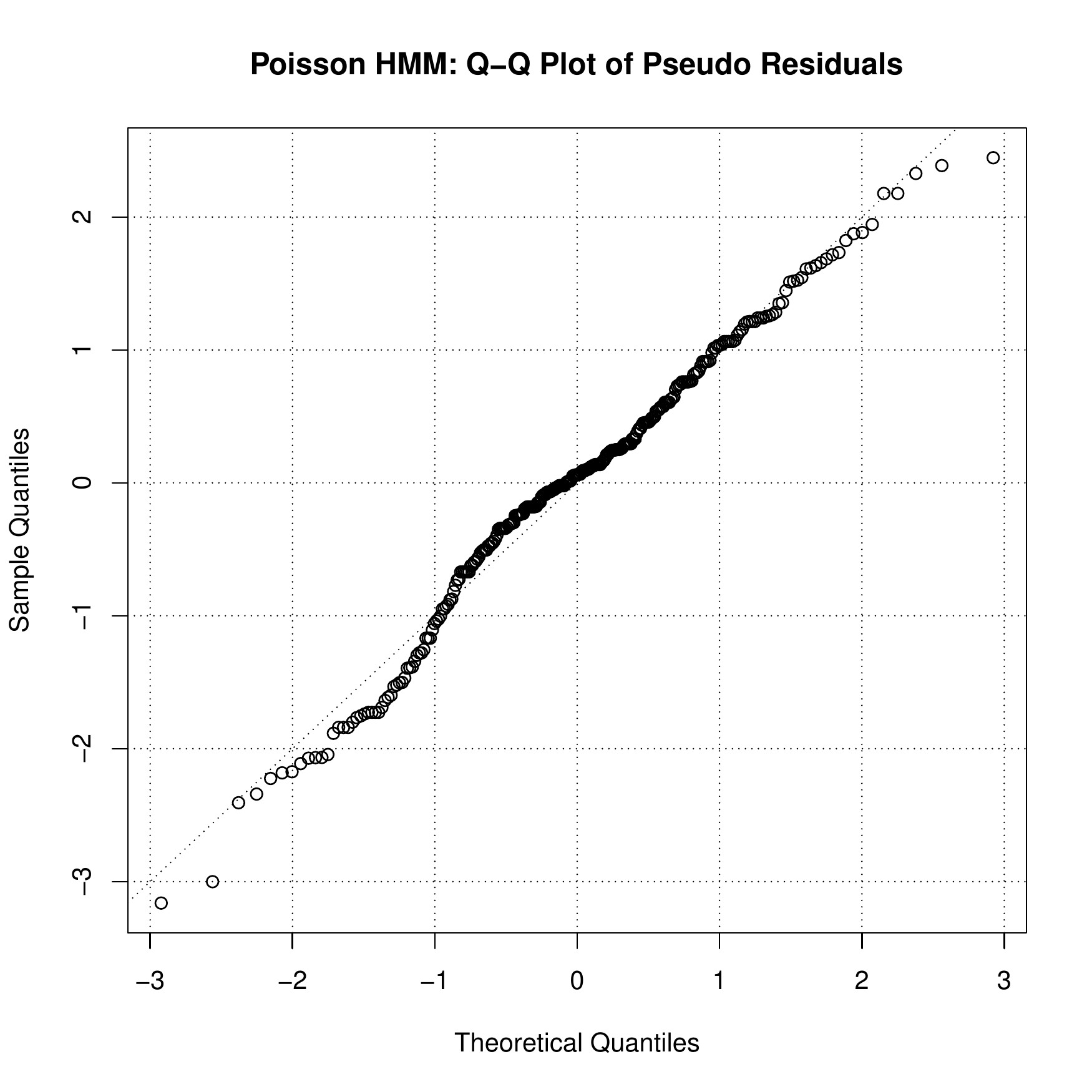}}
  \caption[YouTube Live Channel]{Fig.~\ref{fig:plot:viewers:category:1} shows a plot of viewers of YouTube Live channel for one session. The distribution of the viewers is modelled by a hidden Markov process with state dependent Poisson observation process. The parameters of the model are given by~\cref{eqn:transition:matrix:realdata:study,eqn:observation:matrix:realdata:study}. The QQ-plot for validating the goodness of fit is given in Fig.~\ref{fig:qqplot:youtube}. }
\label{fig:youtube:live} 
\end{figure}

We selected data from the ``entertainment'' category from the YouTube Live dataset. 
The channel contains data from January 01, 2014 to Jan 31, 2014, i.e.\ for a period of one month. 
Fig.~\ref{fig:plot:viewers:category:1} shows the distribution of the viewers of the channel during Jan 01, 2014. 
The parameters of the channel were obtained by the EM-Algorithm~{A.2.3}~in~\cite{zucchini2009hidden}. 
The EM-Algorithm was run for Markov Chain with $2-12$ states. Using the AIC and BIC criteria, we selected that the channel be modelled by a $5$ state Markov chain with the transition matrix in~\eqref{eqn:transition:matrix:realdata:study} and observations following state dependent Poisson distribution with mean given by~\eqref{eqn:observation:matrix:realdata:study}. As can be seen from~\eqref{eqn:transition:matrix:realdata:study} the transition matrix is a first-order Markov chain validating our initial assumptions. Moreover, the transition matrix is diagonally dominant entries ensuring the TP2 assumption. This diagonally domainant entries in the transition matrix in~\eqref{eqn:transition:matrix:realdata:study} models the fact the \veng of the channel changes at a slower time scale compared to the decision epochs (or sampling epochs). The reward depend on both the $g$ in~\eqref{eqn:observation:matrix:realdata:study} and the completion and click rate $\alpha_i$. Since the click rate $\alpha_i$ is not available in the dataset, we assume $\alpha_i = \alpha$. Due to the ordinality of the reward, $\alpha$ is assumed to be equal to $1$. 
The model is validated using the QQ-plot of pseudo-residuals defined in~{Sec.\ {6.1}}~in~\cite{zucchini2009hidden} and is show in Fig.~\ref{fig:qqplot:youtube}. 
As can be seen from Fig.~\ref{fig:qqplot:youtube}, the QQ-plot closely follows the straight line which indicates that the model is a good fit for the data. 

\begin{minipage}{.5\textwidth}
\begin{equation}
	P = 	
	\begin{bmatrix}
		0.94 & 0.06 & 0.00 & 0.00 & 0.00\\
		0.02 & 0.94 & 0.04 & 0.00 & 0.00\\
		0.00 & 0.02 & 0.96 & 0.02 & 0.00\\
		0.00 & 0.00 & 0.06 & 0.91 & 0.03\\
		0.00 & 0.00 & 0.00 & 0.01 & 0.99		
	\end{bmatrix}
	\label{eqn:transition:matrix:realdata:study}
\end{equation}
\end{minipage}%
\begin{minipage}{.5\textwidth}
\begin{equation}
	g = 
	\begin{bmatrix}
		184 & 139 & 102  & 66  & 37
	\end{bmatrix}
	\label{eqn:observation:matrix:realdata:study}
\end{equation}
\end{minipage}

Fig.~\ref{fig:realdata:category:1:compare:schemes} shows the comparison between the various schemes. It can be seen that the policy obtained through Algorithm~\ref{algo:policy:gradient:algorithm} outperforms conventional periodic scheduling by $30\%$.  

\subsubsection{Gaming (Twitch)}
\begin{figure}[!ht]
  \centering
  \subfigure[Plot of the viewers for one session]{\label{fig:plot:viewers:category:2} \includegraphics[scale=0.45]{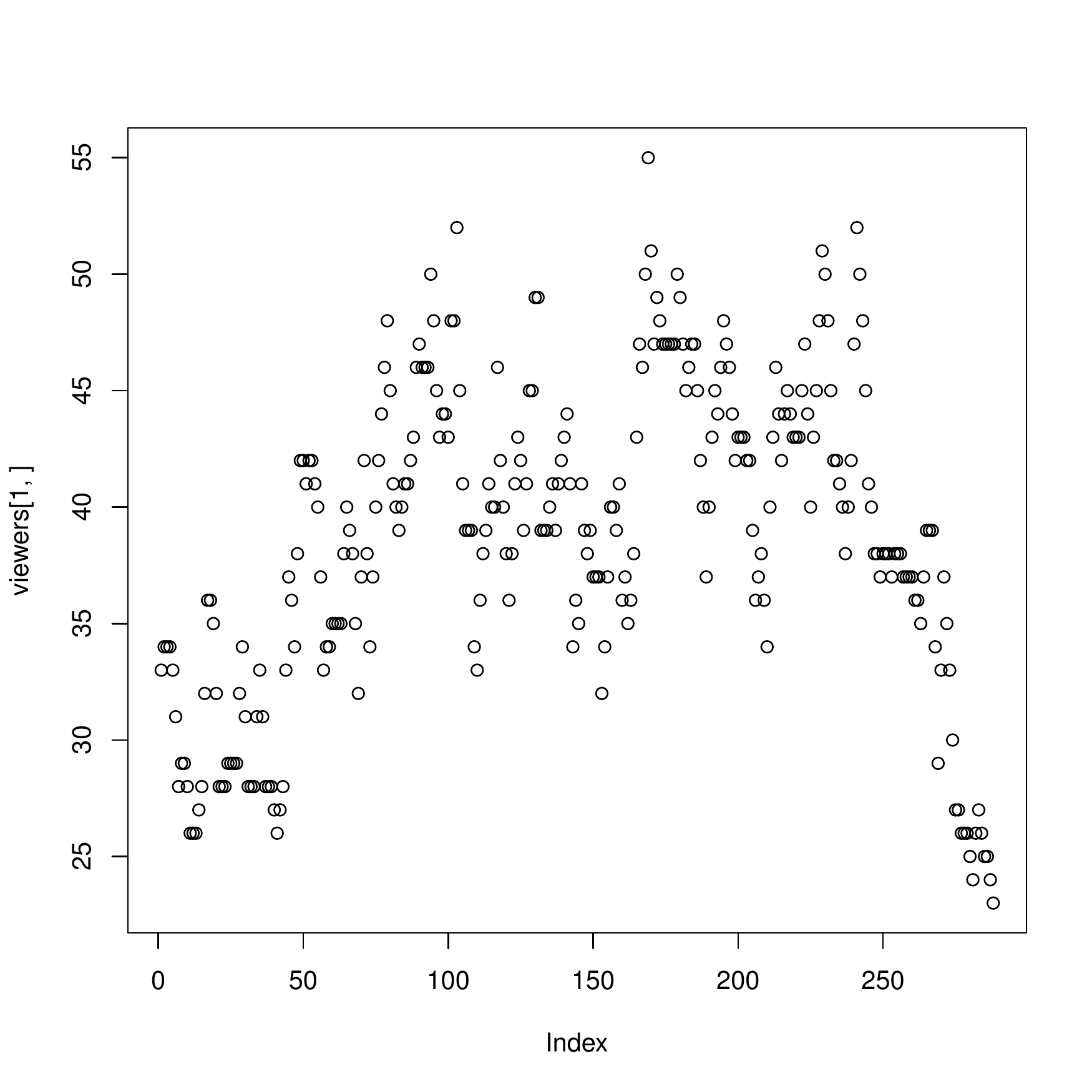}}
  \subfigure[QQ-plot]{\label{fig:qqplot:twitch} \includegraphics[scale=0.45]{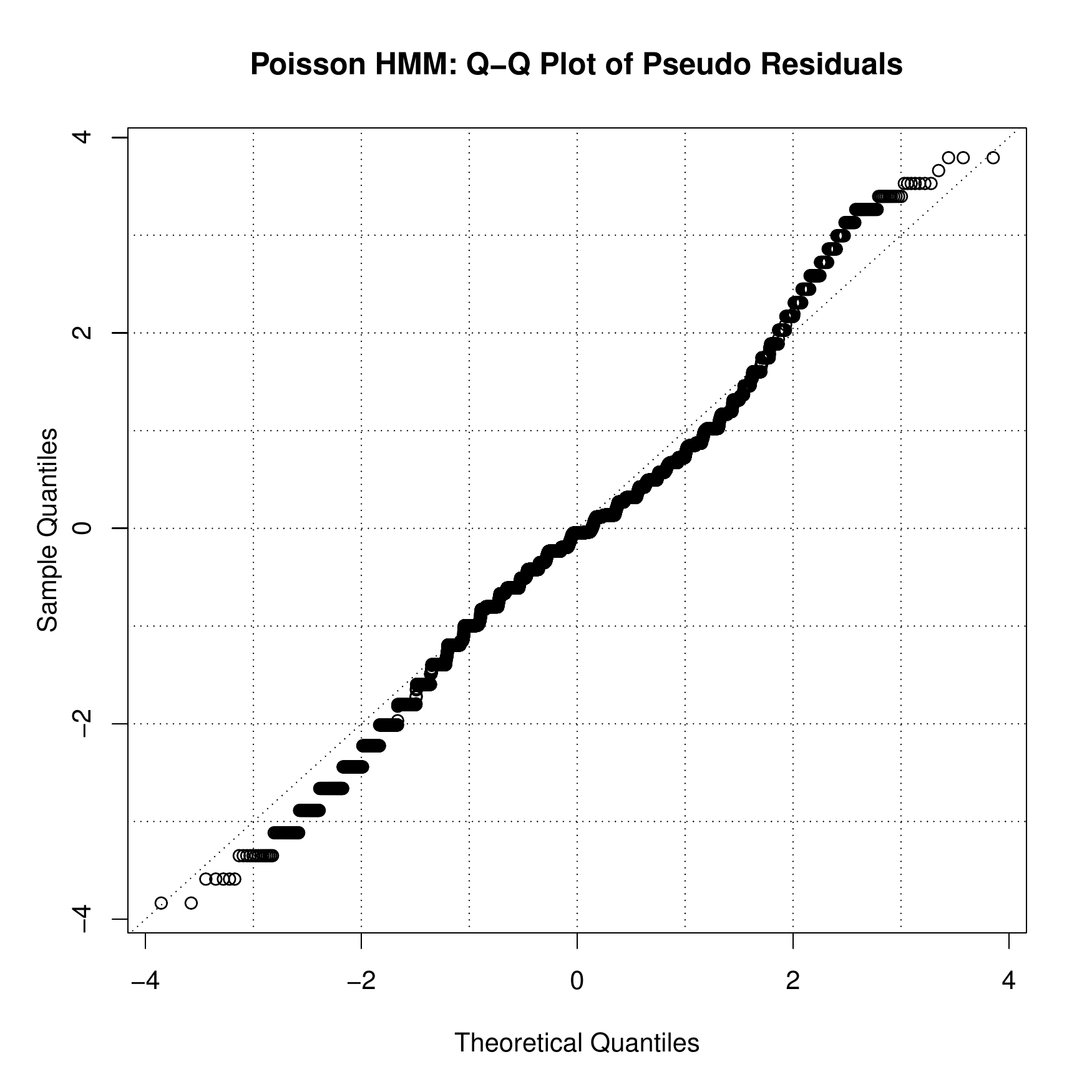}}
  \caption[Twitch Channel]{Fig.~\ref{fig:plot:viewers:category:2} shows a plot of viewers of Twitch channel for one session. The distribution of the viewers is modelled by a hidden Markov process with state dependent Poisson observation process. The parameters of the model are given by~\cref{eqn:transition:matrix:realdata:study:twitch,eqn:observation:matrix:realdata:study:twitch}. The QQ-plot for validating the goodness of fit is given in Fig.~\ref{fig:qqplot:twitch}. }
\label{fig:twitch} 
\end{figure}

The Twitch dataset contains channels with ``gaming'' content. 
Fig.~\ref{fig:plot:viewers:category:2} shows the distribution of the viewers of the channel during Jan 01, 2014. Similar, to the Sec.~\ref{sec:dataset:youtube:live} above, we use the EM-Algorithm in~\cite{zucchini2009hidden} to estimate the parameters of the Markov model. The parameters of the Markov model consisting of the transition matrix and the state dependent mean of the Poisson distribution is as given in~\eqref{eqn:transition:matrix:realdata:study:twitch} and~\eqref{eqn:observation:matrix:realdata:study:twitch}, respectively. The model is validated using QQ-plot of pseudo-residuals and is shown in Fig.~\ref{fig:qqplot:twitch}. 

\begin{minipage}{.40\textwidth}
\begin{equation}
	P = 
	\begin{bmatrix}
0.97& 0.03& 0.00& 0.00& 0.00\\
0.01& 0.96& 0.03& 0.00& 0.00\\
0.00& 0.02& 0.95& 0.03& 0.00\\
0.00& 0.00& 0.02& 0.96& 0.01\\
0.00& 0.00& 0.00& 0.02& 0.98
	\end{bmatrix}
	\label{eqn:transition:matrix:realdata:study:twitch}
\end{equation}
\end{minipage}%
\begin{minipage}{.6\textwidth}
\begin{equation}
	g = 
	\begin{bmatrix}
		55.24& 42.40& 34.65& 28.30& 20.6
	\end{bmatrix}
	\label{eqn:observation:matrix:realdata:study:twitch}
\end{equation}
\end{minipage}

\begin{figure}[!ht]
  \centering
  \subfigure[YouTube Live]{\label{fig:realdata:category:1:compare:schemes} \includegraphics[height=0.25\textheight]{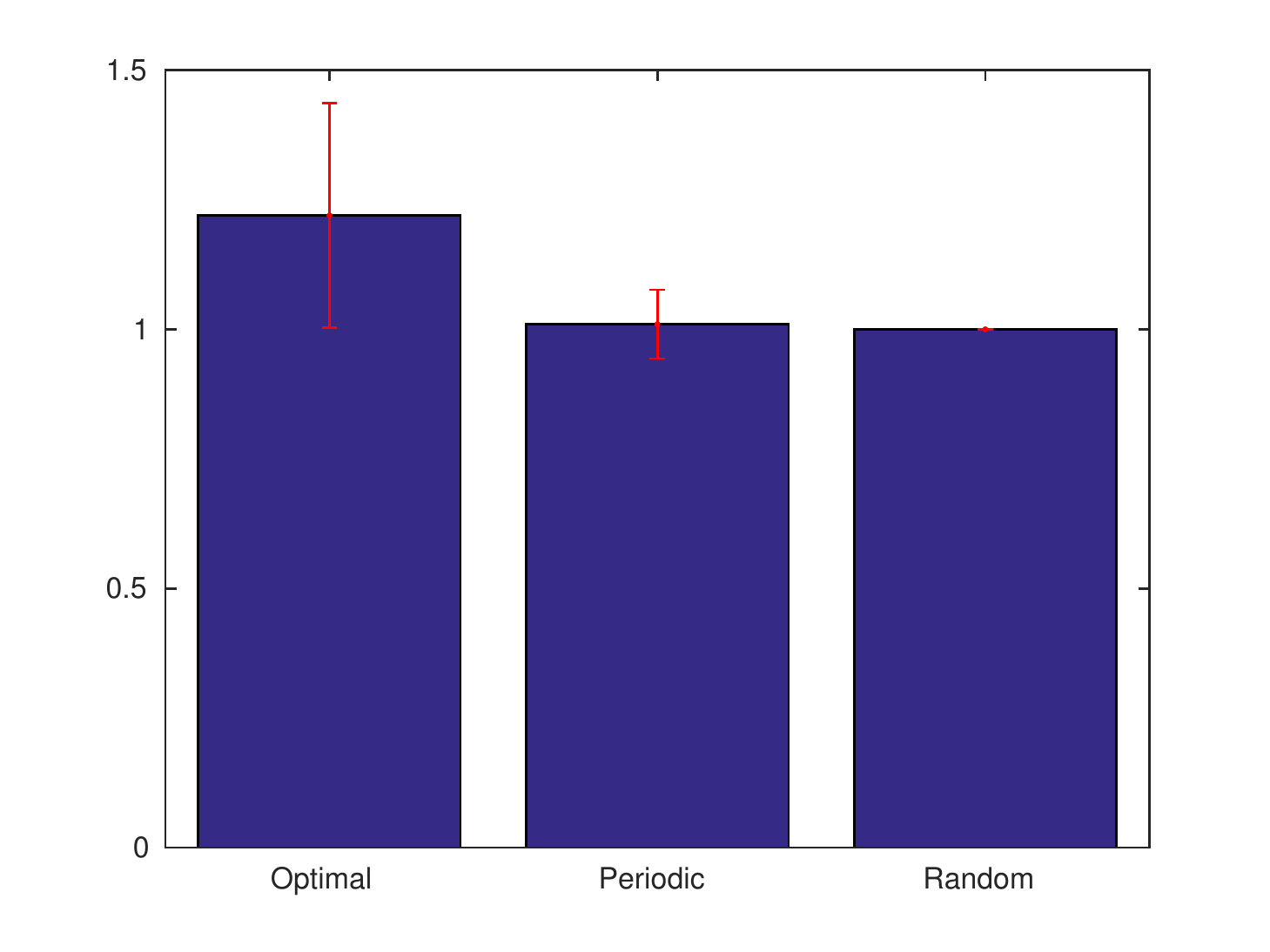}}
  \subfigure[Twitch]{\label{fig:realdata:gaming:compare:schemes} \includegraphics[height=0.25\textheight]{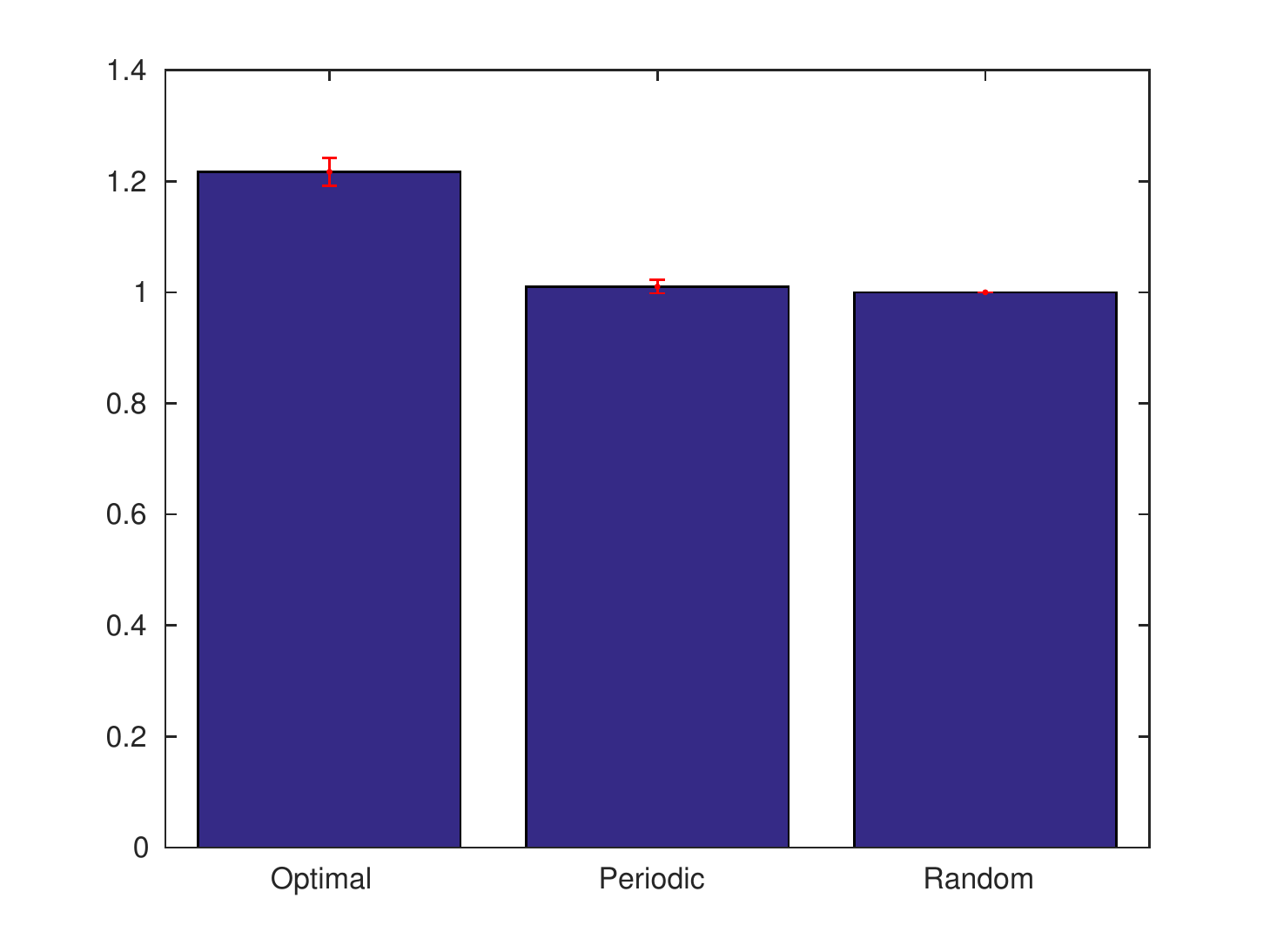}}
  \caption[Comparison of schemes]{Comparison of the various scheduling policies for the YouTube Live (Fig.~\ref{fig:realdata:category:1:compare:schemes}) and Twitch (Fig.~\ref{fig:realdata:gaming:compare:schemes}). The performance of the policy obtained by solving the stochastic approximation algorithm in Algorithm~\ref{algo:policy:gradient:algorithm} is shown as `Optimal'. The policy obtained through Algorithm~\ref{algo:policy:gradient:algorithm} outperforms conventional periodic scheduling (shown as `Periodic') by $20\%$. The random scheduling (shown as `Random') is used for benchmarking. }
\label{fig:compare:schemes:realdata} 
\end{figure}

Fig.~\ref{fig:realdata:gaming:compare:schemes} shows the comparison between the various schemes. Similar to the result in the YouTube Live session, the policy obtained through Algorithm~\ref{algo:policy:gradient:algorithm} outperforms conventional periodic scheduling by close to $20\%$.  
Comparing Fig.~\ref{fig:realdata:category:1:compare:schemes} and Fig.~\ref{fig:realdata:gaming:compare:schemes} we find that the performance of Algorithm~\ref{algo:policy:gradient:algorithm} is lower in the Twitch channel compared to YouTube Live. 
This is due to the fact that Twitch is a subscription based service and the viewers are more ``loyal'' and hence their is less variation in the number of viewers. 
\section{Conclusion}
\label{sec:conclusion}
In this paper, we considered the problem of optimal scheduling of ads on live online social broadcasting channels. 
First, we cast the problem as an optimal multiple stopping problem in the POMDP framework. 
Second, we characterized the structural results of the optimal ad scheduling policy. 
Using the structural results of the optimal ad scheduling policy we computed best approximate policies using stochastic approximation. 
Finally, we validated the results on real datasets. 
First, we illustrated the analysis using synthetic data. In addition, using synthetic data, we showed that the optimal ad scheduling policy outperforms conventional scheduling techniques. 
Second, we show through simulations, that the policy obtained from the multiple stopping problem framework, used for ad scheduling, can be used to detect changes in the ground truth using data from online search. 
Detecting changes in ground truth is useful for optimizing ad strategy. 
Finally, we show through simulations, that the best approximate ad scheduling policies obtained through stochastic approximation outperforms conventional periodic scheduling by $20-30\%$. 

Extension of the current work could involve developing upper and lower myopic bounds to the optimal policy as in~\cite{KP15}, optimizing the ad length and constraints on ad placement. 
These issues promise to offer interesting avenues for future work. 
\appendix
\section{Preliminaries and Definitions}
\label{appendix:def:mlr:lines}
\subsection{First-order stochastic dominance}
\begin{definition}
	Let $\pi_1 \in \Pi$ and $\pi_2 \in \Pi$ be two belief state vectors. 
	Then, $\pi_1$ is greater than $\pi_2$ with respect to first-order stochastic dominance--denoted as $\pi_1 \ge_{s} \pi_2$, if
	\begin{equation}
		\sum_{i = j}^S \pi_1(i) \le \sum_{i=j}^S \pi_2(i) \quad \forall j \in \left\{1,2,\cdots,S\right\}
		\label{eqn:stochastic:dominance:ordering}
	\end{equation}
\end{definition}
\subsection{MLR ordering}
\label{subsec:mlr:ordering}
\begin{definition}
  	\label{def:mlr:ordering}
	Let $\pi_1 \in \Pi$ and $\pi_2 \in \Pi$ be two belief state vectors.  
	Then, $\pi_1$ is greater than $\pi_2$ with respect to Monotone Likelihood Ratio (MLR) ordering--denoted as $\pi_1 \ge_{r} \pi_2$, if
	\begin{equation}
		\pi_1(j) \pi_2(i) \le \pi_2(j)\pi_1(i), \quad i < j, \; i, j \in \left\{1,\dots,S\right\} 
		\label{eqn:mlr:ordering}
	\end{equation}
\end{definition}
MLR ordering over $\Pi$ is a strong condition.  
In order to show threshold structure, we define the following weaker notion of MLR ordering over two types of lines. 
\subsection{MLR ordering over lines}
\label{subsec:mlr:ordering:lines}
First, we define $\mathcal{H}$ as the $S-2$ dimensional linear hyperplane which connects the vertices $e_2,\dots,e_{S}$ as follows: 
\begin{equation}
	\mathcal{H} = \left\{\bar{\pi}:\bar{\pi} \in \mathcal{H} \text{ and } \bar{\pi}(1) = 0\right\}.
	\label{eqn:def:H}
\end{equation}
Figure~\ref{fig:mlr:lines} illustrates the definition~\eqref{eqn:def:H} for an optimal multiple stopping problem with $S=3$. 
Next, we construct two types of lines as follows:
\begin{compactitem}
\item $\mathcal{L}\left(e_1,\bar{\pi}\right)$: For any $\bar{\pi} \in \mathcal{H}$, construct the line $\mathcal{L}(e_1,\bar{\pi})$ that connects $\bar{\pi}$ to $e_1$ as below:
	\begin{equation}
		\mathcal{L}\left(e_1,\bar{\pi}\right) = {\left\{\pi \in \Pi: \pi = \left(1-\gamma\right)\bar{\pi} + \gamma e_1, 0\le \gamma \le 1\right\}}, {\bar{\pi} \in \mathcal{H}}
		\label{eqn:def:L:e1}
	\end{equation}
\item $\mathcal{L}\left(e_S,\bar{\pi}\right)$: For any $\bar{\pi} \in \mathcal{H}$, construct the line $\mathcal{L}(e_S,\bar{\pi})$ that connects $\bar{\pi}$ to $e_S$ as below:
	\begin{equation}
			\mathcal{L}\left(e_S,\bar{\pi}\right) = \left\{\pi \in \Pi: \pi = \left(1-\gamma\right)\bar{\pi} + \gamma e_S, 0\le \gamma \le 1\right\}, 
			{\bar{\pi} \in \mathcal{H}} \label{eqn:def:L:eS}
	\end{equation}
\end{compactitem}
With an abuse of notation, we denote $\mathcal{L}(e_1,\bar{\pi})$ by $\mathcal{L}(e_1)$ and $\mathcal{L}(e_S,\bar{\pi})$ by $\mathcal{L}(e_S)$. Figure~\ref{fig:mlr:lines} illustrates the definition of $\mathcal{L}(e_1)$. 
\begin{definition}[MLR ordering on lines]
	$\pi_1$ is greater than $\pi_2$ with respect to MLR ordering on the lines $\mathcal{L}(e_1)$, denoted as $\pi_1 \ge_{\mathcal{L}_1} \pi_2$, if $\pi_1, \pi_2 \in \mathcal{L}(e_1,\bar{\pi})$, for some $\bar{\pi} \in \mathcal{H}$ and $\pi_1 \ge_{r} \pi_2$. 
\end{definition}
The MLR order is a partial order, however, the MLR ordering on lines is a complete order. The MLR on lines requires less stringent conditions and can be used for devising threshold policies over lines.  

\subsection{TP2 ordering}
\begin{definition}[TP2 ordering]
	A transition probability matrix, $A$ is Totally Positive of order 2 (TP2), if all the second order minors are non-negative i.e.\ the determinants
	\begin{equation}
		\begin{vmatrix}
			a_{i_1j_1} & a_{i_1j_2} \\
			a_{i_2j_1} & a_{i_2j_2}
		\end{vmatrix}
		\ge 0, \forall i_2\ge i_1, j_2\ge j_1
		\label{eqn:definition:TP2}
	\end{equation}
	\label{def:TP2:ordering}
\end{definition}

An important consequence of the TP2 ordering in Definition~\ref{def:TP2:ordering} is the following theorems, which states that the filter $\filter(\belief,y)$ preserves MLR dominance. 
\begin{theorem}[Theorem 10.3.1 in~\cite{krishnamurthy2016partially}]
	If the transition matrix, $P$, and the observation matrix, $B$, satisfies the condition in~\ref{ass:observation} and~\ref{ass:transition}, then 
	\begin{compactitem}
	\item For $\pi_1 \ge_r \pi_2$, the filter satisfies $\filter(\belief_1,y) \ge_r \filter(\belief_2,y)$. 
	\item For $\pi_1 \ge_r \pi_2$, $\sigma(\pi_1,y) \ge_s \sigma(\pi_2,y)$ 
	\end{compactitem}
	\label{thm:VK:filter}
\end{theorem}

\begin{theorem}[\cite{kiessler2005comparison}]
	$\pi_1 \ge_s \pi_2$ if and only if for any increasing function $\phi(\cdot)$, $\E_{\pi_1}\left\{\phi(x)\right\} \ge \E_{\pi_2}\left\{\phi(x)\right\}$
	\label{thm:VK:Value}
\end{theorem}
\section{Proof of Prop.~\ref{prop:V:increase:pi}, Prop.~\ref{prop:W:decrease:l}, Prop.~\ref{prop:nested}, Theorem~\ref{thm:main}}
\label{appendix:proofs}
To prove Prop.~\ref{prop:V:increase:pi}, Prop.~\ref{prop:W:decrease:l} and Prop.~\ref{prop:nested}, we assume that the proposition hold for all values less than $k$. 
\subsection{Proof of Prop.~\ref{prop:V:increase:pi}}
Recall from~\eqref{eqn:dyn}, 
\begin{equation*}
	V_{k}(\pi,l) = \underset{u \in \{1,2\}}{\text{max}}  Q_{k}(\pi,l,u), 
\end{equation*}
To prove Prop.~\ref{prop:V:increase:pi}, we show $Q_{k}(\pi,l,u)$ is increasing in $\pi$ for $u=\left\{1,2\right\}$. 

Recall from~\eqref{eqn:Q:stop}, 
\begin{equation*}
	\scalebox{0.94}[1]{$Q_{k}(\pi,l,1) = \reward{\pi} + \rho \sum_{y} V_{k-1}(\Tpiy,l-1)\Spiy$}, 
\end{equation*}
Using Theorem~\ref{thm:VK:filter}, Theorem~\ref{thm:VK:Value} and the induction hypothesis, the term $\sum_{y} V_{k-1}(\Tpiy,l-1)\Spiy$ is increasing in $\pi$. From Assumption~\ref{ass:u}, $\reward{\pi}$ is increasing in $\pi$. The proof for $Q_{k}(\pi,l,2)$ increasing in $\pi$ is similar and is omitted. Hence, $V_{k}(\pi,l)$ is increasing in $\pi$. \IEEEQEDhere

\subsection{Proof of Prop.~\ref{prop:W:decrease:l}}
  The proof follows by induction. 
  Recall from~\eqref{w_k_l}, we have
	  \begin{equation}
		  \scalebox{0.90}[1]{$W_{k}(\pi,l-1) = 	\sum_{y} {W_{k-1}(\Tpiy,l-1)\Spiy \mathcal{I}_{C_{k}^{l-1}}(\pi)} + 
	  		 	{\reward{\pi} \mathcal{I}_{C_{k}^{l-2}\cap S_{k}^{l-1}}(\pi)} + 
				{\sum_{y} W_{k-1}(\Tpiy,l-2)\Spiy \mathcal{I}_{S_{k}^{l-2}}(\pi)}$}
    \label{w_k_lm1}
  	\end{equation}
  Hence, we compare $W_{k}(\pi,l)$ and $W_{k}(\pi,l-1)$ in the following $4$ regions:
	\begin{compactitem}
	\item[a.)] $S_{k}^{l-2}: $
	    \begin{equation*}
		    W_{k}(\pi,l) - W_{k}(\pi,l-1) = \sum_{y} ( W_{k-1}(\Tpiy,l-1) - W_{k-1}(\Tpiy,l-2) )\Spiy,
	    \end{equation*}
	    which is non-negative by the induction assumption. 
    \item[b.)] 	 $C_{k}^{l-2} \cap S_{k}^{l-1}: $  
	    \begin{equation*}
		    W_{k}(\pi,l) - W_{k}(\pi,l-1) = \sum_{y} W_{k-1}(\Tpiy,l-1) \Spiy  - \reward{\pi},
	    \end{equation*}
	    which is non-negative since $\scalebox{0.9}[1]{$\pi \in S_k^{l-1}$}$. 
	  \item[c.)] $C_{k}^{l-1} \cap S_{k}^{l}: $ 
	    \begin{equation*}
		    W_{k}(\pi,l) - W_{k}(\pi,l-1) = \reward{\pi} - \sum_{y} W_{k-1}(\Tpiy,l-1) \Spiy,
	    \end{equation*}
	    which is non-negative since $\scalebox{0.9}[1]{$\pi \in C_k^{l-1}$}$. 
	  \item[d.)] $C_{k}^{l}: $ 
	    \begin{equation*}
		    W_{k}(\pi,l) - W_{k}(\pi,l-1) = \sum_{y} ( W_{k-1}(\Tpiy,l) - W_{k-1}(\Tpiy,l-1) )\Spiy,
	    \end{equation*}
	    which is non-negative by the induction assumption. 
	\end{compactitem}  
\IEEEQEDhere
\subsection{Proof of Prop.~\ref{prop:nested}}
If $\pi \in S_{k}^{l-1}$, then $\reward{\pi} \ge \sum_{y} W_{k-1}(\Tpiy,l-1)\Spiy$. By Prop.~\ref{prop:W:decrease:l}, $\reward{\pi} \ge \sum_{y} W_{k-1}(\Tpiy,l)\Spiy$. Hence $\pi \in S_{k}^l$. \IEEEQEDhere

\subsection{Proof of Theorem~\ref{thm:main}}
\textbf{Existence of optimal policy:} 
In order to show the existence of a threshold policy of $\mathcal{L}_1$, we need to show that $Q_{k+1}(\pi,l,2) - Q_{k+1}(\pi,l,1)$ is supermodular in $\pi \in \mathcal{L}(e_1,\bar{\pi})$. 
Since, 
\begin{equation*}
	Q_{k+1}(\pi,l,2) - Q_{k+1}(\pi,l,1) =  \rho \sum_{y} W_{k}(\Tpiy,l)\Spiy  - \reward{\pi}.
\end{equation*}
We need to show that $\rho \sum_{y} W_{k}(\Tpiy,l)\Spiy  - \reward{\pi}$ is decreasing in $\pi$. 
\begin{align}
	 \rho \sum_{y} W_{k}(\Tpiy,l)\Spiy  - \reward{\pi}  
	& = \sum_{y} \left( \rho W_{k}(\Tpiy,l)  - \reward{\pi} \right) \Spiy \nonumber \\
	& = \sum_{y} \left( \left(\rho W_{k}(\Tpiy,l)-\rho \reward{\Tpiy}\right) - \left(\reward{\pi} -\rho \reward{\Tpiy}\right) \right) \Spiy \nonumber \\
	& = \rho \sum_{y} \left(W_{k}(\Tpiy,l)-\reward{\Tpiy}\right) \Spiy - r^\prime (I - \rho P^\prime) \pi  \label{eqn:sufficent:W:decreasing}
\end{align}
The term $r^\prime (I - \rho P^\prime) \pi$ in~\eqref{eqn:sufficent:W:decreasing} is decreasing in $\pi$ due to our assumption. 
Hence, to show that $\rho \sum_{y} W_{k}(\Tpiy,l)\Spiy  - \reward{\pi}$ is decreasing in $\pi$ it is sufficent to show that $W_{k}(\pi,l)  - \reward{\pi}$ is decreasing in $\pi$. 
Define, 
\begin{equation}
	\bar{W}_{k}(\pi,l) \triangleq  W_{k}(\pi,l)  - \reward{\pi}
	\label{eqn:define:W:bar}
\end{equation}

Now, $\bar{W}_{k}(\pi,l) = $  
\begin{align}
	& \scalebox{0.90}[1]{$\left(\sum_{y} \rho \left( \left(\bar{W}_{k-1}(\Tpiy,l) + \reward{\Tpiy} \right) - \reward{\pi}\right) \Spiy \right) \mathcal{I}_{C_{k}^{l}}(\pi)  + 
	\left(\sum_{y} \rho \left( \left(\bar{W}_{k-1}(\Tpiy,l-1) + \reward{\Tpiy} \right) - \reward{\pi}\right) \Spiy \right) \mathcal{I}_{S_{k}^{l}}(\pi) $}\nonumber \\
	& = \scalebox{0.90}[1]{$\left(\sum_{y} \left(\rho \bar{W}_{k-1}(\Tpiy,l) \Spiy \right) -r^\prime (I -\rho P)^\prime \pi \right)  \mathcal{I}_{C_{k}^{l}}(\pi)  + 
	\left(\sum_{y} \left(\rho \bar{W}_{k-1}(\Tpiy,l-1)\Spiy \right) -r^\prime (I -\rho P)^\prime \pi \right)  \mathcal{I}_{S_{k}^{l}}(\pi)$} \label{eqn:W:bar:inductive:step}
\end{align}
We prove using induction that $\bar{W}_{k}(\pi,l)$ is decreasing in $\pi$, using the recursive relation over $k$ in~\eqref{eqn:W:bar:inductive:step}. 

For $k=0$,
\begin{equation}
	\bar{W}_{0}(\pi,l) = W_{0}(\pi,l)-r^\prime \pi 
	 = V_{0}(\pi,l)-V_{0}(\pi,l-1)-r^\prime \pi 
	 \label{eqn:W:0:pi:l}
\end{equation}
The initial conditions of the value iteration algorithm can be chosen such that $\bar{W}_{0}(\pi,l)$ in~\eqref{eqn:W:0:pi:l} is decreasing in $\pi$. A suitable choice of the initial conditions is given below: 
\begin{equation}
	V_{0}(\pi,l) = r^\prime \left(\sum_{j=0}^{l-1} \rho^j P^j\right)^\prime \pi.
	\label{eqn:value:iteration:initial:conditions}
\end{equation}
The intuition behind the initial conditions in~\eqref{eqn:value:iteration:initial:conditions} is that the value function, $V_{0}(\pi,l)$ gives the expected total reward if we stop $l$ times successively starting at belief $\pi$. 
Hence, it is clear that $\bar{W}_{k}(\pi,l)$ is decreasing in $\pi$, if $\bar{W}_{k-1}(\pi,l)$ is decreasing in $\pi$, finishing the induction step.  

\textbf{Characterization of the switching curve $\Gamma_l$:} For each $\bar{\pi} \in \mathcal{H}$ construct the line segment $\mathcal{L}(e_1,\bar{\pi})$. 
The line segment can be described as $(1-\varepsilon) \bar{\pi} + \varepsilon e_1$. On the line segment $\mathcal{L}(e_1,\bar{\pi})$ all the belief states are MLR orderable. 
Since $\mu^*(\pi,l)$ is monotone decreasing in $\pi$, for each $l$, we pick the largest $\varepsilon$ such that $\mu^*(\pi,l)=1$. The belief state, $\pi^{\varepsilon^*, \bar{\pi}}$ is the threshold belief state, where $\varepsilon^* = \text{inf }\{\varepsilon \in [0,1]: \mu^*({\pi^{\varepsilon,\bar{\pi}}}) = 1\}$. Denote by $\Gamma(\bar{\pi}) = \pi^{\varepsilon^*, \bar{\pi}}$. 
The above construction implies that there is a unique threshold $\Gamma(\bar{\pi})$ on $\mathcal{L}(e_1,\bar{\pi})$. The entire simplex can be covered by considering all pairs of lines $\mathcal{L}(e_1,\bar{\pi})$, for $\bar{\pi} \in \mathcal{H}$, i.e. $\Pi(X) = \cup_{\bar{\pi} \in \mathcal{H}} \mathcal{L}(e_1,\bar{\pi})$. Combining, all points yield a unique threshold curve in $\Pi(X)$ given by $\Gamma = \cup_{\bar{\pi} \in \mathcal{H}} \Gamma(\bar{\pi})$. 

\textbf{Connectedness of $S^l$:} Since $e_1 \in S^l$ for all $l$, call $S_a^l$, the subset of $S^l$ that contains $e_1$. 
Suppose $S_b^l$ is the subset that was disconnected from $S_a^l$. 
Since every point on $\Pi(X)$ lies on the line segment $\mathcal{L}(e_1,\bar{\pi})$, for some $\bar{\pi}$, there exists a line segment starting from $e_1 \in S^l_a$ that would leave the region $S_a^l$, pass through the region where action $2$ is optimal and then intersect region $S_b^l$, where action $1$ is optimal. 
But, this violates the requirement that the policy $\mu^*(\pi,l)$ is monotone on $\mathcal{L}(e_1,\bar{\pi})$. 
Hence, $S_a^l$ and $S_b^l$ are connected. 

\textbf{Connectedness of $C^l$:} Assume $e_X \in C^l$, otherwise $C^l = \phi$ and there is nothing to prove. 
Call the region that contains $e_X$ as $C^l_a$. 
Suppose $C_b^l \subset C^l$ is disconnected from $C_a^l$. 
Since every point in $\Pi(X)$ can be covered by a line segment $\mathcal{L}(e_X,\bar{\pi})$, for some $\bar{\pi}$.   
Then, there exists a line starting from $e_X \in C^l_a$ would leave region $C_a^l$, pass through the region where action $1$ is optimal and then intersect the region $C_b^l$ (where action $2$ is optimal). But this violates the monotone property of $\mu^*(\pi,l)$.  

\textbf{Sub-setting structure:} The proof is straightforward from Prop.~\ref{prop:nested}. \IEEEQEDhere

\subsection{Proof of Theorem~\ref{thm:constraints:parameter}}
\label{appendix:proof:thm:constraints:parameter}
For $l_1 > l_2$, due to the sub-setting structure in Theorem~\ref{thm:main} $S^{l_2} \subset S^{l_1}$. 
This implies the following
\begin{align}
	\mu_\theta(l_2,\pi) &\ge \mu_\theta(l_1,\pi) \nonumber \\
	\begin{bmatrix} 0& 1& \theta_{l_2} \end{bmatrix} \begin{bmatrix} \pi \\ -1 \end{bmatrix} &\ge \begin{bmatrix} 0& 1& \theta_{l_1} \end{bmatrix} \begin{bmatrix} \pi \\ -1 \end{bmatrix} \nonumber \\
	\begin{bmatrix} 0& 0& \theta_{l_2}-\theta_{l_1} \end{bmatrix} \begin{bmatrix} \pi \\ -1 \end{bmatrix} &\ge 0 \label{eqn:final:step:proof:subsetting:structure}
\end{align}
It is straightforward to check that the conditions in~\eqref{eqn:cond:parameter} in Theorem~\ref{thm:constraints:parameter} satisfy the conditions in~\eqref{eqn:final:step:proof:subsetting:structure}. \IEEEQEDhere 


\bibliographystyle{IEEEtran}

\begin{thebibliography}{10}
\providecommand{\url}[1]{#1}
\csname url@samestyle\endcsname
\providecommand{\newblock}{\relax}
\providecommand{\bibinfo}[2]{#2}
\providecommand{\BIBentrySTDinterwordspacing}{\spaceskip=0pt\relax}
\providecommand{\BIBentryALTinterwordstretchfactor}{4}
\providecommand{\BIBentryALTinterwordspacing}{\spaceskip=\fontdimen2\font plus
\BIBentryALTinterwordstretchfactor\fontdimen3\font minus
  \fontdimen4\font\relax}
\providecommand{\BIBforeignlanguage}[2]{{%
\expandafter\ifx\csname l@#1\endcsname\relax
\else
\language=\csname l@#1\endcsname
\fi
#2}}
\providecommand{\BIBdecl}{\relax}
\BIBdecl

\bibitem{SOW13}
T.~Smith, M.~Obrist, and P.~Wright, ``Live-streaming changes the (video)
  game,'' in \emph{Proc.\ of the 11th European Conference on Interactive TV and
  Video}.\hskip 1em plus 0.5em minus 0.4em\relax ACM, 2013, pp. 131--138.

\bibitem{BBM04}
S.~Bollapragada, M.~R. Bussieck, and S.~Mallik, ``Scheduling commercial
  videotapes in broadcast television,'' \emph{Oper. Res.}, vol.~52, no.~5, pp.
  679--689, Oct. 2004.

\bibitem{PC15}
D.~G. Popescu and P.~Crama, ``Ad revenue optimization in live broadcasting,''
  \emph{Management Science}, vol.~62, no.~4, pp. 1145--1164, 2015.

\bibitem{SSS15}
S.~Seshadri, S.~Subramanian, and S.~Souyris, ``Scheduling spots on
  television,'' 2015.

\bibitem{KM11}
H.~Kang and M.~P. McAllister, ``Selling you and your clicks: examining the
  audience commodification of google,'' \emph{Journal for a Global Sustainable
  Information Society}, vol.~9, no.~2, pp. 141--153, 2011.

\bibitem{TC13}
R.~Terlutter and M.~L. Capella, ``The gamification of advertising: analysis and
  research directions of in-game advertising, advergames, and advertising in
  social network games,'' \emph{Journal of Advertising}, vol.~42, no. 2-3, pp.
  95--112, 2013.

\bibitem{TST11}
J.~Turner, A.~Scheller-Wolf, and S.~Tayur, ``Scheduling of dynamic in-game
  advertising,'' \emph{Operations Research}, vol.~59, no.~1, pp. 1--16, 2011.

\bibitem{AMM12}
N.~Archak, V.~Mirrokni, and S.~Muthukrishnan, ``Budget optimization for online
  campaigns with positive carryover effects,'' in \emph{Proc. of the 8th
  International Conference on Internet and Network Economics}.\hskip 1em plus
  0.5em minus 0.4em\relax Springer-Verlag, 2012, pp. 86--99.

\bibitem{AMM10}
N.~Archak, V.~S. Mirrokni, and S.~Muthukrishnan, ``Mining advertiser-specific
  user behavior using adfactors,'' in \emph{Proceedings of the 19th
  International Conference on World Wide Web}, ser. WWW '10.\hskip 1em plus
  0.5em minus 0.4em\relax ACM, 2010, pp. 31--40.

\bibitem{nak85}
T.~Nakai, ``The problem of optimal stopping in a partially observable markov
  chain,'' \emph{Journal of optimization Theory and Applications}, vol.~45,
  no.~3, pp. 425--442, 1985.

\bibitem{sta87}
W.~Stadje, ``An optimal k-stopping problem for the poisson process,'' in
  \emph{Mathematical Statistics and Probability Theory}.\hskip 1em plus 0.5em
  minus 0.4em\relax Springer, 1987, pp. 231--244.

\bibitem{Nik99}
M.~Nikolaev, ``On optimal multiple stopping of markov sequences,'' \emph{Theory
  of Probability \& Its Applications}, vol.~43, no.~2, pp. 298--306, 1999.

\bibitem{Ann15}
A.~Krasnosielska-Kobos, ``Multiple-stopping problems with random horizon,''
  \emph{Optimization}, vol.~64, no.~7, pp. 1625--1645, 2015.

\bibitem{bertsekas1995dynamic}
D.~P. Bertsekas, D.~P. Bertsekas, D.~P. Bertsekas, and D.~P. Bertsekas,
  \emph{Dynamic programming and optimal control}.\hskip 1em plus 0.5em minus
  0.4em\relax Athena Scientific Belmont, MA, 1995, vol.~1, no.~2.

\bibitem{ER15}
E.~Bayraktar and R.~Kravitz, ``Quickest detection with discretely controlled
  observations,'' \emph{Sequential Analysis}, vol.~34, no.~1, pp. 77--133,
  2015.

\bibitem{GBL14}
J.~Geng, E.~Bayraktar, and L.~Lai, ``Bayesian quickest change-point detection
  with sampling right constraints,'' \emph{IEEE Transactions on Information
  Theory}, vol.~60, no.~10, pp. 6474--6490, 2014.

\bibitem{Lai97}
T.~L. Lai, ``On optimal stopping problems in sequential hypothesis testing,''
  \emph{Statistica Sinica}, vol.~7, no.~1, pp. 33--51, 1997.

\bibitem{Lai01}
------, \emph{Sequential analysis}.\hskip 1em plus 0.5em minus 0.4em\relax
  Wiley Online Library, 2001.

\bibitem{NJ10}
S.~H.~J. Alexander G.~Nikolaev, ``Stochastic sequential decision-making with a
  random number of jobs,'' \emph{Operations Research}, vol.~58, no.~4, pp.
  1023--1027, 2010.

\bibitem{ST05}
S.~Savin and C.~Terwiesch, ``Optimal product launch times in a duopoly:
  Balancing life-cycle revenues with product cost,'' \emph{Operations
  Research}, vol.~53, no.~1, pp. 26--47, 2005.

\bibitem{LPVZ15}
I.~Lobel, J.~Patel, G.~Vulcano, and J.~Zhang, ``Optimizing product launches in
  the presence of strategic consumers,'' \emph{Management Science}, vol.~62,
  no.~6, pp. 1778--1799, 2015.

\bibitem{WRA11}
K.~E. Wilson, R.~Szechtman, and M.~P. Atkinson, ``A sequential perspective on
  searching for static targets,'' \emph{European Journal of Operational
  Research}, vol. 215, no.~1, pp. 218 -- 226, 2011.

\bibitem{AKL16}
M.~Atkinson, M.~Kress, and R.-J. Lange, ``When is information sufficient for
  action? search with unreliable yet informative intelligence,''
  \emph{Operations Research}, vol.~64, no.~2, pp. 315--328, 2016.

\bibitem{Ask07}
I.~D. Askwith, ``Television 2.0: Reconceptualizing tv as an engagement
  medium,'' Ph.D. dissertation, Massachusetts Institute of Technology, 2007.

\bibitem{HZZZ06}
H.~Yu, D.~Zheng, B.~Y. Zhao, and W.~Zheng, ``Understanding user behavior in
  large-scale video-on-demand systems,'' \emph{SIGOPS Oper. Syst. Rev.},
  vol.~40, no.~4, pp. 333--344, Apr. 2006.

\bibitem{KD07}
V.~Krishnamurthy and D.~V. Djonin, ``Structured threshold policies for dynamic
  sensor scheduling--a partially observed markov decision process approach,''
  \emph{IEEE Transactions on Signal Processing}, vol.~55, no.~10, pp.
  4938--4957, Oct 2007.

\bibitem{krishnamurthy2016partially}
V.~Krishnamurthy, \emph{Partially Observed Markov Decision Processes}.\hskip
  1em plus 0.5em minus 0.4em\relax Cambridge University Press, 2016.

\bibitem{Kri13}
------, ``How to schedule measurements of a noisy {M}arkov chain in decision
  making?'' \emph{IEEE Transactions on Information Theory}, vol.~59, no.~7, pp.
  4440--4461, July 2013.

\bibitem{Kri11}
------, ``Bayesian sequential detection with phase-distributed change time and
  nonlinear penalty -- {A} {POMDP} {L}attice programming approach,'' \emph{IEEE
  Transactions on Information Theory}, vol.~57, no.~10, pp. 7096--7124, Oct
  2011.

\bibitem{10.2307/2583338}
J.~J. M.~Eric~Johnson, ``Infinitesimal perturbation analysis: A tool for
  simulation,'' \emph{The Journal of the Operational Research Society},
  vol.~40, no.~3, pp. 243--254, 1989.

\bibitem{pflug2012optimization}
G.~C. Pflug, \emph{Optimization of stochastic models: the interface between
  simulation and optimization}.\hskip 1em plus 0.5em minus 0.4em\relax Springer
  Science \& Business Media, 2012, vol. 373.

\bibitem{spall2005introduction}
J.~C. Spall, \emph{Introduction to stochastic search and optimization:
  estimation, simulation, and control}.\hskip 1em plus 0.5em minus 0.4em\relax
  John Wiley \& Sons, 2005, vol.~65.

\bibitem{wang2008stochastic}
I.-J. Wang and J.~C. Spall, ``Stochastic optimisation with inequality
  constraints using simultaneous perturbations and penalty functions,''
  \emph{International Journal of Control}, vol.~81, no.~8, pp. 1232--1238,
  2008.

\bibitem{JWCZ13}
M.~Joo, K.~C. Wilbur, B.~Cowgill, and Y.~Zhu, ``Television advertising and
  online search,'' \emph{Management Science}, vol.~60, no.~1, pp. 56--73, 2013.

\bibitem{GMPBSMB09}
J.~Ginsberg, M.~H. Mohebbi, R.~S. Patel, L.~Brammer, M.~S. Smolinski, and
  L.~Brilliant, ``Detecting influenza epidemics using search engine query
  data,'' \emph{Nature}, vol. 457, no. 7232, pp. 1012--1014, 2009.

\bibitem{Pires:2015}
K.~Pires and G.~Simon, ``Youtube live and twitch: A tour of user-generated live
  streaming systems,'' in \emph{Proceedings of the 6th ACM Multimedia Systems
  Conference}, ser. MMSys '15.\hskip 1em plus 0.5em minus 0.4em\relax ACM,
  2015, pp. 225--230.

\bibitem{zucchini2009hidden}
W.~Zucchini and I.~L. MacDonald, \emph{Hidden Markov models for time series: an
  introduction using R}.\hskip 1em plus 0.5em minus 0.4em\relax CRC press,
  2009.

\bibitem{KP15}
V.~Krishnamurthy and U.~Pareek, ``Myopic bounds for optimal policy of {POMDPs}:
  An extension of {L}ovejoy's structural results,'' \emph{Operations Research},
  vol.~62, no.~2, pp. 428--434, 2015.

\bibitem{kiessler2005comparison}
P.~C. Kiessler, ``Comparison methods for stochastic models and risks,''
  \emph{Journal of the American Statistical Association}, vol. 100, no. 470,
  pp. 704--704, 2005.

\end{thebibliography}

\end{document}